\documentclass[preprint,12pt]{aastex}
\usepackage{graphicx}

\newcommand{\etal}{et al.}

\begin{document}

\title{THE ROLE OF ELECTRON CAPTURES IN CHANDRASEKHAR MASS MODELS FOR TYPE IA
SUPERNOVAE}

\author{
Franziska Brachwitz\altaffilmark{1}, 
David J. Dean\altaffilmark{2},
W. Raphael Hix\altaffilmark{2,3}, 
Koichi Iwamoto\altaffilmark{4}, 
Karlheinz Langanke\altaffilmark{6},
Gabriel Mart\'{\i}nez-Pinedo\altaffilmark{6},
Ken'ichi Nomoto\altaffilmark{5,7}, 
Michael R.~Strayer\altaffilmark{2},
Friedrich-K. Thielemann\altaffilmark{1,2},
Hideyuki Umeda\altaffilmark{7}
}
\altaffiltext{1}{Department of Physics and Astronomy, University of Basel,
CH-4056 Basel, Switzerland}
\altaffiltext{2}{Physics Division, Oak Ridge National Laboratory, Oak Ridge, 
TN 37831-4576, USA}
\altaffiltext{3}{Department of Physics and Astronomy, University of Tennessee,
Knoxville, TN 37996-1200, USA}  
\altaffiltext{4}{Department of Physics, Nihon University, Tokyo 101-8308, Japan}
\altaffiltext{5}{Department of Astronomy, University of Tokyo, Tokyo 113-0033, 
Japan}
\altaffiltext{6}{Institute of Physics \& Astronomy, University of Aarhus,
DK-8000 Aarhus C, Denmark}
\altaffiltext{7}{Research Center for the Early Universe, School of 
Science, University of Tokyo, Tokyo 113-0033, Japan}

\begin{abstract}
The Chandrasekhar mass model for Type Ia Supernovae (SNe Ia) has received
increasing support from recent comparisons of observations with light curve 
predictions and modeling of synthetic spectra. It explains SN Ia events via  
thermonuclear explosions of accreting white dwarfs in binary stellar systems, 
being caused by central carbon ignition when the white dwarf approaches the 
Chandrasekhar mass.
As the electron gas in white dwarfs is degenerate, characterized by high 
Fermi energies for the high density regions in the center, electron capture on 
intermediate mass and Fe-group nuclei plays an important role in explosive 
burning. Electron capture affects the central electron fraction $Y_e$, 
which determines the composition of the ejecta from such explosions.
Up to the present, astrophysical tabulations based on shell model matrix 
elements were only available for light nuclei in the sd-shell.
Recently new Shell Model Monte Carlo (SMMC) and large-scale
shell model diagonalization calculations have also been performed for
pf-shell nuclei. These lead in general to a reduction of electron capture
rates in comparison with previous, more phenomenological, approaches.
Making use of these new shell model based rates,
we present the first results for the composition of Fe-group nuclei produced in 
the central regions of SNe Ia and possible changes in the constraints on model 
parameters
like ignition densities $\rho_{ign}$ and burning front speeds $v_{def}$.
\end{abstract}


\section{Introduction}
Electron capture is an important phenomenon in the late phases
of stellar evolution, during stellar collapse, and in explosive events
like Type I supernovae (SNe Ia), Type II supernovae (SNe II), and possibly
in X-ray bursts (rp-process). Its takes place in high density matter, where 
the Fermi energy of a degenerate
electron gas is sufficiently large to overcome the energy thresholds 
given by the negative Q-values of such
reactions. The Fermi energy exceeds a fraction of an MeV 
in late burning stages, allowing electron capture initially only for a few 
selected nuclei like $^{33}$S and  $^{35}$Cl (with small energy thresholds of 
0.247~MeV or $^{35}$Cl, respectively), 
and goes beyond an MeV during Fe-core collapse or SNe Ia explosions.

In the following we discuss the effects of electron capture on nuclei
during the flame propagation in SNe Ia.
There are strong observational and theoretical indications that
SNe Ia (a classification based on the absence of hydrogen lines and
the presence of a specific SiII line in their spectra)
are thermonuclear explosions of accreting white dwarfs
\citep[e.g.,][]{nomoto94,wheeler95,hoefkhok96,nomiwa97,nomoto97c,hoeflich97,nugent97,hoefwhth98,branch98}
with high accretion rates, which permit relatively stable H- and
He-shell burning and lead to a growing C/O white dwarf. When 
the white dwarf mass grows close to the Chandrasekhar mass, contraction sets 
in and the central density becomes high enough to
ignite carbon fusion under degenerate conditions.
The environment of a degenerate electron gas provides a pressure which
depends only on the density. Therefore, the initial heat generation
does not lead to pressure increase and expansion, which would result in
controlled and stable burning. Instead, a thermonuclear runaway occurs. The 
burning front propagates through the whole star, causing complete disruption 
without a remnant.

The high Fermi energy of the degenerate electron gas in the 
white dwarf leads to efficient electron capture in the high density burning 
regions and reduces $Y_e=<Z/A>$, the electron fraction or equivalently
the average proton to nucleon ratio, during explosive burning in the center.
This is an important factor, controlling the isotopic composition ejected from 
such explosions (i.e., how neutron-rich is the matter produced).
If the central density exceeds a critical value, electron capture can cause
a dramatic reduction in the pressure of degenerate electrons and can therefore
induce collapse (``accretion induced collapse", AIC) of the white dwarf
\citep{noko91}. Thus, 
electron capture on intermediate mass and Fe-group nuclei plays a crucial role
for the burning front propagation in SNe Ia.
When $Y_e$'s are attained which correspond to the $Z/A$ ratios
of nuclei more neutron-rich than stability, the reverse beta-decays are 
also relevant.

Weak interactions describing electron/positron capture or beta-decay are
either Fermi or Gamow-Teller transitions. Fermi transitions are only 
populating the isobaric analog state. Gamow-Teller transitions are
distributed over many final states, with the maximum strength centered
around the Gamow-Teller giant resonance.
Folding these distributions with the thermal energy distribution of electrons
(and the thermal population of target states) gives Gamow-Teller transitions
the dominant role in burning at high temperatures and densities. Thus, in 
order to unravel the dynamics of the burning front propagation in SNe Ia, it is
important to have an understanding of the Gamow-Teller strength functions in
both the electron-capture and beta-decay reactions for unstable
pf-shell nuclei. Up to present, astrophysical tabulations based on shell model
matrix elements have been only available for nuclei in the sd-shell (A=17-40)
\citep{ffn80}. For heavier nuclei, more simplified approaches, based on average
positions of the Gamow-Teller giant resonance and average matrix elements, were
applied by the same authors for nuclei up to A=60, supplemented by existing
experimental information \citep[][hereinafter
denoted FFN]{ffn82,ffn85}. Revisions for sd-shell nuclei with accurate shell 
model wave functions
and experimental transition strengths where available were calculated 
\citep{takahara89,oda94}, but for pf-shell nuclei the more appropriate
shell model methods are only now becoming available. \citet{aufderheide94}
performed a detailed study, in order to understand which nuclei are of primary
importance for a variety of densities and $Y_e$ values.
They found $^{55-68}$Co, $^{56-69}$Ni, $^{53-62}$Fe,
$^{53-63}$Mn, $^{64-74}$Cu, $^{49-54}$Sc, $^{50-58}$V, $^{52-59}$Cr,
$^{49-54}$Ti, $^{74-80}$Ga, $^{77-80}$Ge, $^{83}$Se, $^{80-83}$As, and
$^{75}$Zn to be important. These nuclei have recently been addressed with 
Quasi-particle Random Phase
Approxomation (QRPA) methods \citep{klapdor99}, but more importantly, shell
model diagonalization and shell model Monte Carlo approaches have also recently
become available, which allow sufficiently accurate calculations in the
pf-shell and at finite temperatures \citep{koonin97}.
First applications seem to reproduce the measured
GT-distributions well and differ significantly from FFN
 \citep{dean98,caur99a,caur99b}.

The unfortunate situation is that until present there exist two sources of
uncertainties in SNe Ia, related to either (i) the nuclear physics input 
discussed above or (ii) astrophysical modeling of the central ignition density
$\rho_{ign}$ and the flame propagation speed $v_{def}$ in connection with 
hydrodynamic instabilities. Clear constraints for the latter can only be
obtained when former is known with high accuracy. Though the white dwarf models
can successfully account for the basic observational features of SNe Ia, the
exact binary evolution that leads to SNe Ia has not yet been identified 
\citep[see, however,][]{hachisu99a,hachisu99b}.  High accretion rates onto the 
white dwarf progenitor cause a stronger heating and thus a higher central 
temperature and pressure, which
favor earlier ignition at lower densities $\rho_{ign}$.
Carbon fusion apparently starts with a deflagration, i.e. a subsonic burning 
front \citep{nomthiyok84}.
The propagation of the burning front occurs initially via heat conduction
in the degenerate electron gas (with a burning front thickness
of the order $10^{-4}$-$10^{-5}$cm).
Instabilities of various scales lead to burning front propagation
via convection and can accelerate the effective burning front speed.
Multi-dimensional hydrodynamical simulations of the
flame propagation have been attempted by several groups, though the
results are still preliminary 
\citep{livne93,arnliv94,khokhlov95,niemeyer95,niewoos97,hilnie97}. These simulations have
suggested that a carbon deflagration wave might initially propagate at a speed
$v_{\rm def}$ as slow as a few percent of the sound speed $v_{\rm s}$
in the central region of the white dwarf.  For example, \citet{niemeyer95} 
obtained $v_{\rm def}/v_{\rm s} \sim$ 0.015.
After an initial deflagration in the central layers, the
deflagration can turn into a detonation at lower densities
$\rho_{tr}$ \citep{khokhlov91,woosweav94,niemeyer99}.

When summarizing the preceding discussion, we not notice that present modeling 
uncertainties in type Ia supernovae are related to ignition densities 
$\rho_{ign}$ (progenitor evolution), the treatment of the burning front 
propagation with hydrodynamic instabilities ($v_{def}$ and $\rho_{tr}$), and 
nuclear uncertainties for the electron capture rates on Fe-group
nuclei in this high density and temperature environment. 
At temperatures exceeding $5\times 10^9$K, nuclear reactions involving
strong plus electromagnetic forces are fast enough to attain a chemical 
equilibrium (in this context also referred to as Nuclear Statistical
Equilibrium, NSE) which determines nuclear abundances
on timescales much shorter than the typical burning front propagation times
of the order of seconds. This reduces abundance uncertainties from the
knowledge of reaction cross sections to those
of masses and partition functions \citep{clayton83}. 
But weak interactions act on longer
timescales and their uncertainties enter directly.
SNe Ia are the main producers of Fe-peak elements in the 
Galaxy \citep[see e.g. the discussion in][]{iwamoto99} . Electron capture on 
nuclei in the incinerated material is
responsible for the total neutron to proton ratio of matter and thus are
crucial to the isotopic composition of Fe-group nuclei.
The amount of electron capture depends on both $v_{\rm def}$ and
$\rho_{ign}$. The central density of the white dwarf during ignition 
$\rho_{ign}$
affects the electron chemical potential. The burning front speed $v_{def}$
affects the time duration of matter at high temperatures, and with
it the availability of free protons which can experience electron capture and 
the shape of the high energy tail of the electron energy distributions. 
Thus, the existing constraints for the production of neutron-rich Fe-group 
nuclei in SNe Ia can be translated into constraints for
these parameters describing the burning front propagation.

In a recent paper we \citep{iwamoto99} have attempted to find such constraints
for $\rho_{ign}$, $v_{def}$ and $\rho_{tr}$ via comparison with the
solar Fe-group composition and chemical evolution models, but were still using
the FFN electron capture rates. In the present paper we will test how
dependent the conclusions are on variations and improvements to the
set of weak interactions employed.

\section {Weak Interactions in the Fe-Group}
\newcommand{\ttbs}{\char'134}
\newcommand{\AmS}{{\protect\the\textfont2
  A\kern-.1667em\lower.5ex\hbox{M}\kern-.125emS}}

The systematic study of stellar weak interaction rates in the mass range
of concern here ($A=45-60$) was pioneered by Fuller, Fowler and
Newman in a series of papers 
in the early eighties \citep{ffn80,ffn82,ffn85}. 
These authors noticed the
extraordinary role played by the Gamow-Teller (GT) giant resonance for 
stellar electron capture and, more strikingly, also for beta-decay.
Unlike in the laboratory, $\beta$-decay
under stellar conditions is significantly increased due to thermal
population of the GT back resonances in the parent nucleus; the GT back
resonances are the states reached by the strong GT transitions in the
inverse process (electron capture) built on the ground and excited
states \citep{ffn80,ffn82,ffn85},
allowing for a transition with a large nuclear matrix element and
increased phase space. Indeed, Fuller, Fowler and Newman concluded that the
$\beta$-decay rates under collapse conditions are dominated by the decay
of the back resonance.
The relevant momentum transfers in type Ia supernovae is low enough to
safely neglect the contributions of forbidden transitions to the weak
interaction rates. Furthermore, the $Y_e$ values encountered in type Ia
supernovae stay large enough to involve only nuclei in NSE for which the GT
transition is not Pauli-blocked; the latter is expected to happen for nuclei
at the neutron shell closure $N=40$ \citep{Pauli}.

The GT contribution to the electron capture and $\beta$-decay rates
has been parametrized by FFN on the basis of the independent particle
model. To complete the FFN rate estimate, the GT contributions were
supplemented by a contribution simulating low-lying transitions, using
experimental data wherever possible.
The FFN rates were updated and extended to heavier nuclei
by \citet{aufderheide94}. 
In recent years, however, the parametrization of the GT contribution, as
adopted in FFN, became questionable when comparisons were made to
upcoming experimental information about the GT distribution in pf-shell nuclei.
These  data clearly indicate that the GT
strength is both quenched and fragmented over several
states at modest excitation energies in the daughter nucleus
\citep{gtdata1,gtdata4,gtdata2,gtdata3,gtdata5}. Thus the need
for an improved theoretical description was soon realized
\citep{Aufderheide91,Aufderheide93a,Aufderheide93b}. It also became 
apparent that a reliable reproduction of the GT distribution in nuclei
requires large shell model calculations which account for all correlations
among the valence nucleons in a major oscillator shell.

Such calculations in a complete major shell (usually referred to as
$0 \hbar \omega$ shell model calculations) are now possible 
using the recently developed Shell Model Monte Carlo (SMMC) 
\citep{Johnson92,lang93,alhassid94,dean95}. Rather than solving the 
many-body problem by diagonalization of the Hamiltonian $H$, the SMMC method
describes the nucleus by a canonical ensemble at temperature  
$T=\beta^{-1}$ and employs a Hubbard-Stratonovich linearization 
\citep{Hubbard,Stratonovich57} of the  
imaginary-time many-body propagator, $e^{-\beta H}$, to express  
observables as path integrals of one-body propagators in fluctuating  
auxiliary fields \citep{lang93}. Since Monte 
Carlo techniques  
avoid an explicit enumeration of the many-body states, they can be  
used in model spaces far larger than those accessible to conventional  
methods. The Monte Carlo results are in principle exact and in  
practice are subject only to controllable sampling and discretization  
errors. A comprehensive review of the SMMC method, a detailed
description of the underlying ideas, its formulation,
and numerical realization can be found in \citet{koonin97}.

\citet{Langanke95} reported on the first complete $pf$-shell
calculation of nuclei in the mass range $A=50-62$. Most important for the
present context, it reproduced all experimentally available 
total Gamow-Teller strengths GT$_+$ well, that is after scaling the spin
operator with the universal quenching factor which accounts for the
contribution of intruder states from outside the model space and appears
to be $A$-independent within the pf-shell.
[The GT$_+$ transitions describe the direction where a proton is changed
into a neutron, like in electron capture or $\beta^+$-decay.]
However, stellar weak interaction rates have a strong phase space
dependence and hence are more sensitive to the GT strength distribution
in nuclei than to the total strength. The calculation of strength
distributions within the SMMC method is in principal possible, but it
involves a numerical inverse Laplace transform which is notoriously
difficult to perform. Nevertheless  \citet{radha97}
succeeded in extracting SMMC GT$_+$ strength
distributions for nuclei in the mass range $A=50-64$ 
and again the agreement with data has been quite satisfactory.
However, it became already apparent in \citet{radha97} that the
SMMC model yields only an ``averaged'' GT strength distribution, as the
statistical noise inherent in the Monte Carlo data allows only to
determine the first moments of the distribution. Thus, the total strength,
centroid and width are well reproduced via the inverse
Laplace transform, but weak transitions to individual states
outside the GT centroid distribution could not be resolved.

Motivated by the successful reproduction of all experimental GT$_+$ strength
distributions, the SMMC approach has subsequently been used to calculate
stellar electron capture rates for several nuclei in the mass range
$A=50-64$ \citep{dean98}. That study included those nine nuclei for which 
experimental data
are available ($^{54,56}$Fe, $^{58,60,62,64}$Ni, $^{51}$V, $^{55}$Mn,
$^{59}$Co), but it also predicted rates for other nuclei
of interest for supernovae ($^{45}$Sc, $^{55,57}$Co, $^{56}$Ni, $^{50,52}$Cr,
$^{55,58}$Fe, and $^{50}$Ti). We note that these calculations were
the first which considered the complete $0 \hbar \omega$ model
space, and they also were consistently performed at finite temperature.
The latter issue had been circumvented in earlier studies which assumed
that the GT strength distributions on excited states are identical to
the one built on the ground state, only shifted upwards in energy 
by the excitation energy of the parent state \citep[FFN,][]{aufderheide94}.

When compared to the FFN parametrization of the GT centroids, the SMMC
calculations showed some systematic deviations. Firstly, for even-even
parent nuclei the GT$_+$ strength generally peaks at lower excitation
energies in the daughter than was assumed by FFN. As a consequence 
one would intuitively expect the SMMC rates to be larger than the FFN
rates for electron capture on even-even nuclei. However, they turned out to
be approximately the same, since FFN often intuitively
compensated for the smaller GT contribution (due to the shift in centroid) 
by an added low-lying transition strength. Secondly, for odd-$A$
nuclei FFN have placed the GT centroid at significantly lower energies
than found in the SMMC results and in the data. Consequently for these
nuclei the GT contribution to the electron capture rate has been
noticeably overestimated in FFN. Moreover for many odd-$A$ nuclei the GT
resonance part in the FFN parametrization dominates the capture rates,
with the added low-lying strength component being rather unimportant.
The SMMC calculations, on the other hand, indicate that the GT
contribution to the rate should be reduced by nearly two orders of
magnitude, making the rate sensitive to the weak transitions at low
excitation energies in the daughter nucleus. This is a rather non-trivial
situation for the SMMC approach, since these weak
components in the GT distribution at low excitation energies are 
difficult to resolve, as mentioned above.

As the SMMC calculations suggest significant modifications of the
stellar electron capture rates, we are motivated to investigate
potential effects of these modifications on the dynamics and the
nucleosynthesis. Unfortunately the SMMC rates available are not complete
\citep{dean98}, as 
beta-decay rates are missing and the capture rates on odd-$A$ nuclei should be
supplemented by the contributions to low-lying states.
These shortcomings can be overcome in large-scale shell model diagonalization
calculations. These approaches have recently made significant progress
and, combined with improved computer technologies, currently allow
for diagonalization
in model spaces large enough (involving typically 10 million or more
configurations) to guarantee that the GT
strength distribution is reasonably converged \citep{caur99a}.
Nevertheless shell model diagonalization is still a computationally 
formidable task and a complete compilation of weak stellar
interaction rates, although possible, is a rather time-consuming
project.

Recently stellar electron and
beta-decay rates have been calculated by shell model diagonalization
for several key nuclei \citep{Martinez99,lanpi98,lanpi99} and
confirm the trend already observed in the SMMC
studies. Systematic deviations from the GT parametrization assumed in
the FFN compilation lead to significantly smaller electron capture rates on
odd-$A$ nuclei and odd-odd nuclei 
(even if the weak low-energy components are properly
included). 
The diagonalization calculations yield, on average,  slightly smaller 
capture rates on even-even nuclei than the FFN and the SMMC approaches. The
latter is due to the fact that
the diagonalization studies employed
a slightly improved version of the residual interaction used in
\citet{dean98}. This correction removed the slight overestimation in the
shell gap at the nucleon number $N=28$, discussed in \citet{Langanke95}.
The modified interaction reproduces quite nicely the ground state and 
prolate deformed bands in $^{56}$Ni \citep{rudolph99}.

What consequences do the misplacement of the GT centroids have for the
competing $\beta$ decays? In odd-A and even-even nuclei (the daughters
of electron capture on odd-odd nuclei) experimental 
data and shell model studies
place the back-resonance at higher excitation energies than assumed by
FFN. Correspondingly, its
population becomes less likely at temperatures prevailing in a SN Ia.
Hence the contribution of the back-resonance
to the $\beta$ decay rates for even-even and odd-A nuclei decreases. 
In contrast, the shell model $\beta$ decay rate for odd-odd nuclei
often are slightly larger than the FFN rates, because
for these nuclei all available data, 
and all shell model calculations
indicate that the back-resonance resides actually at lower excitation
energies than previously parametrized. 

To incorporate the modifications of the FFN rates, as suggested by the large
shell model diagonalization studies, we follow the approximate procedure
suggested in \citet{Martinez99}. These authors compared the FFN  and shell
model rates for several typical 
nuclei and, based on this comparison, suggested to multiply the FFN
electron capture rates by 0.2 (for even-even nuclei), 0.1 (odd-$A$), and
0.04 (odd-odd), while the FFN beta-decay rates might be scaled by 0.05
(even-even), 0.025 (odd-$A$) and 1.7 (odd-odd). In the following we will 
use these simple scalings to simulate potential modifications of the
rates not provided by the SMMC approach. We emphasize here, however,
that these scalings (denoted in the further discussion SMFA) are rather crude, 
and in principle are different for
each nucleus and depend on temperature and density.

\begin{figure}[hpt]
\begin{center}
    \includegraphics[width=4cm,angle=90]{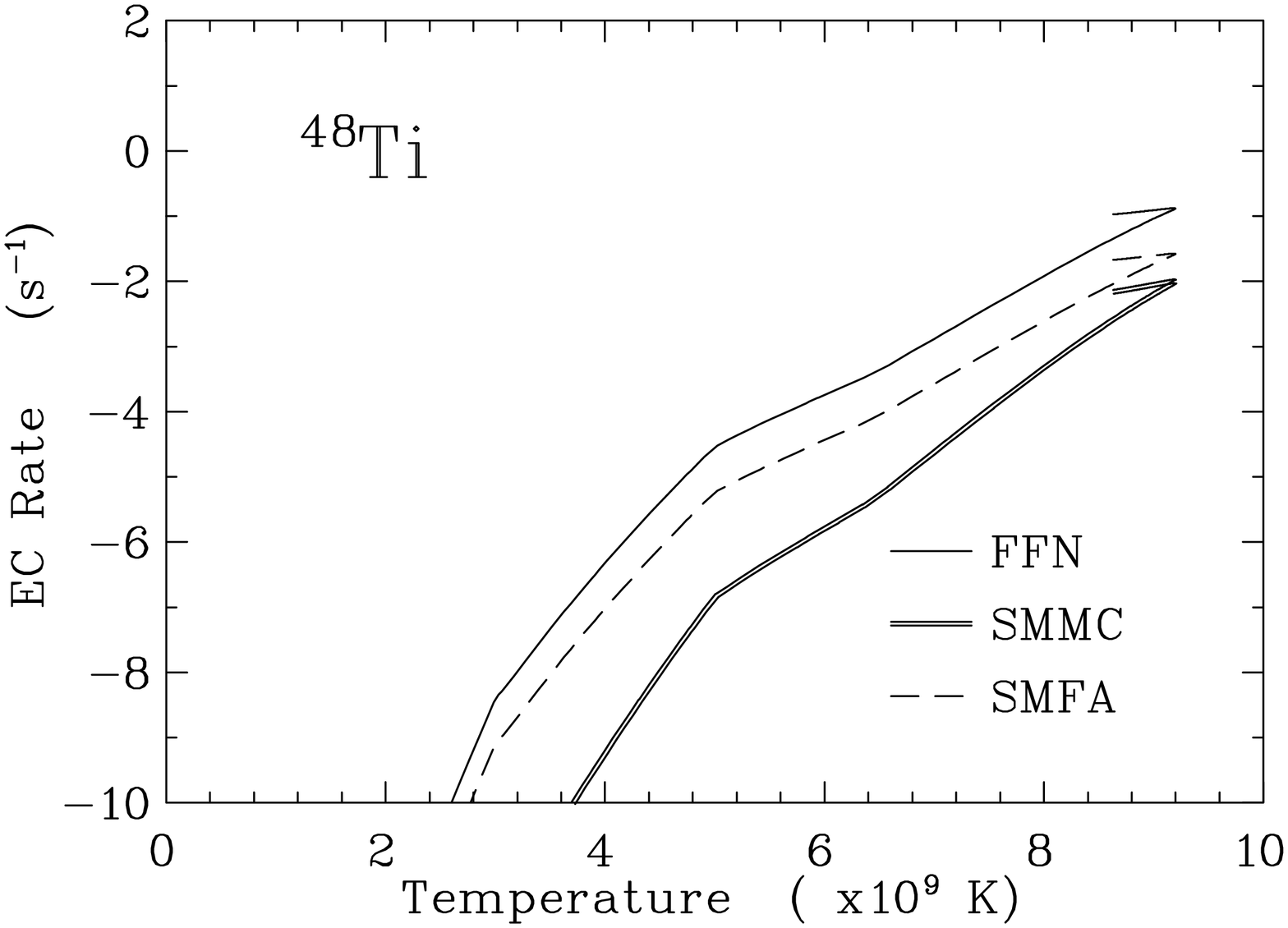}
    \includegraphics[width=4cm,angle=90]{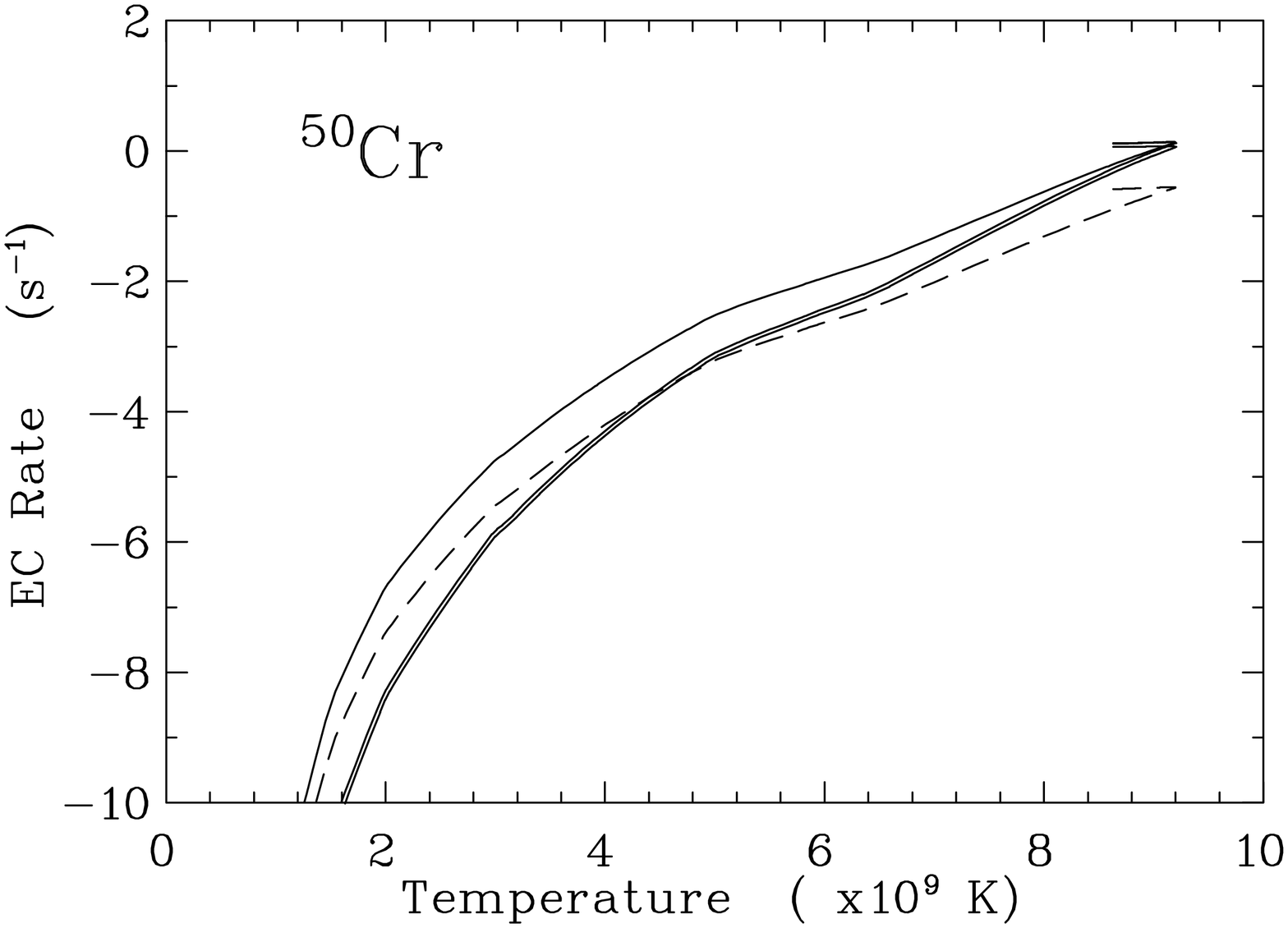}
    \includegraphics[width=4cm,angle=90]{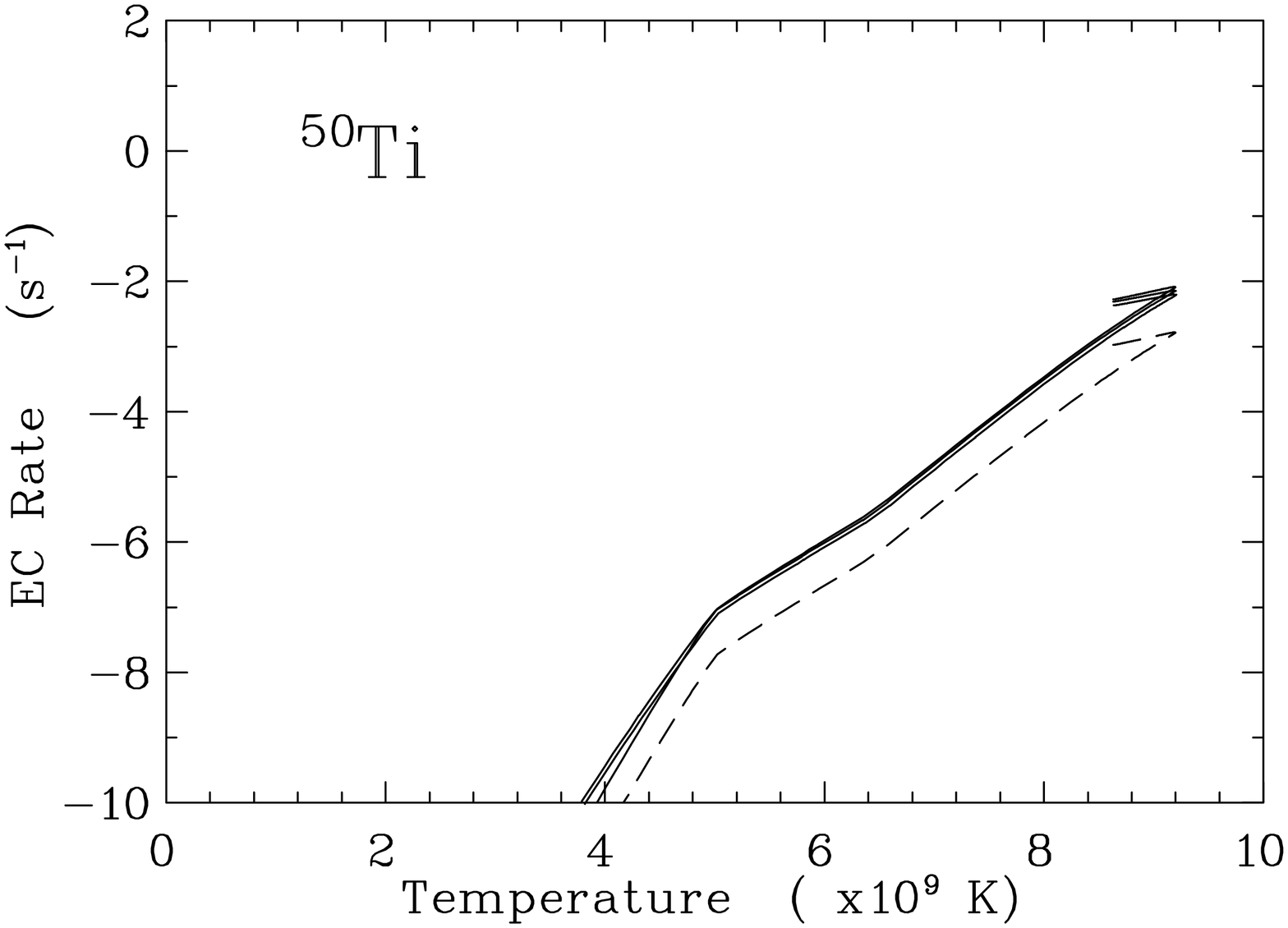}
    \includegraphics[width=4cm,angle=90]{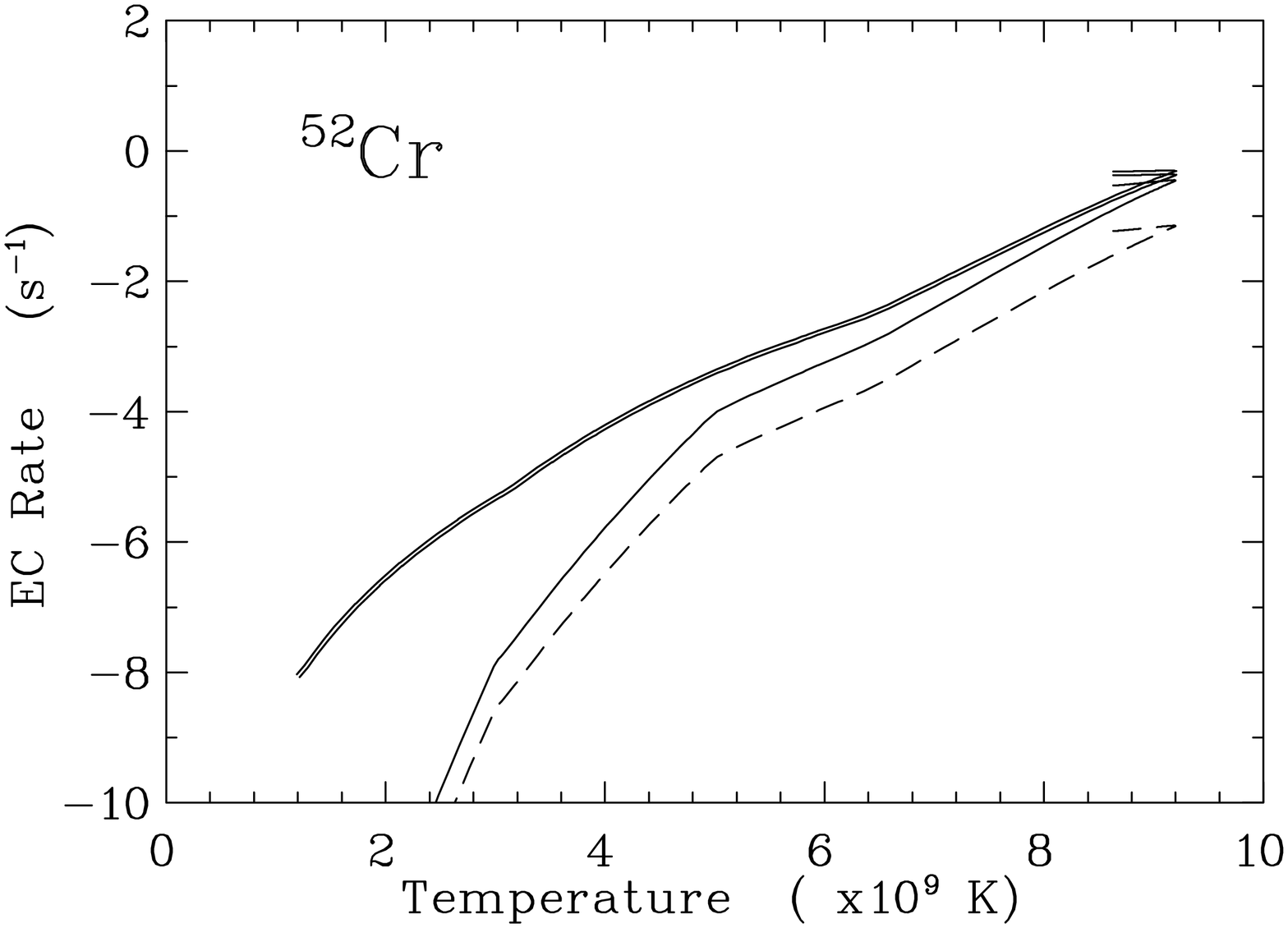}
    \includegraphics[width=4cm,angle=90]{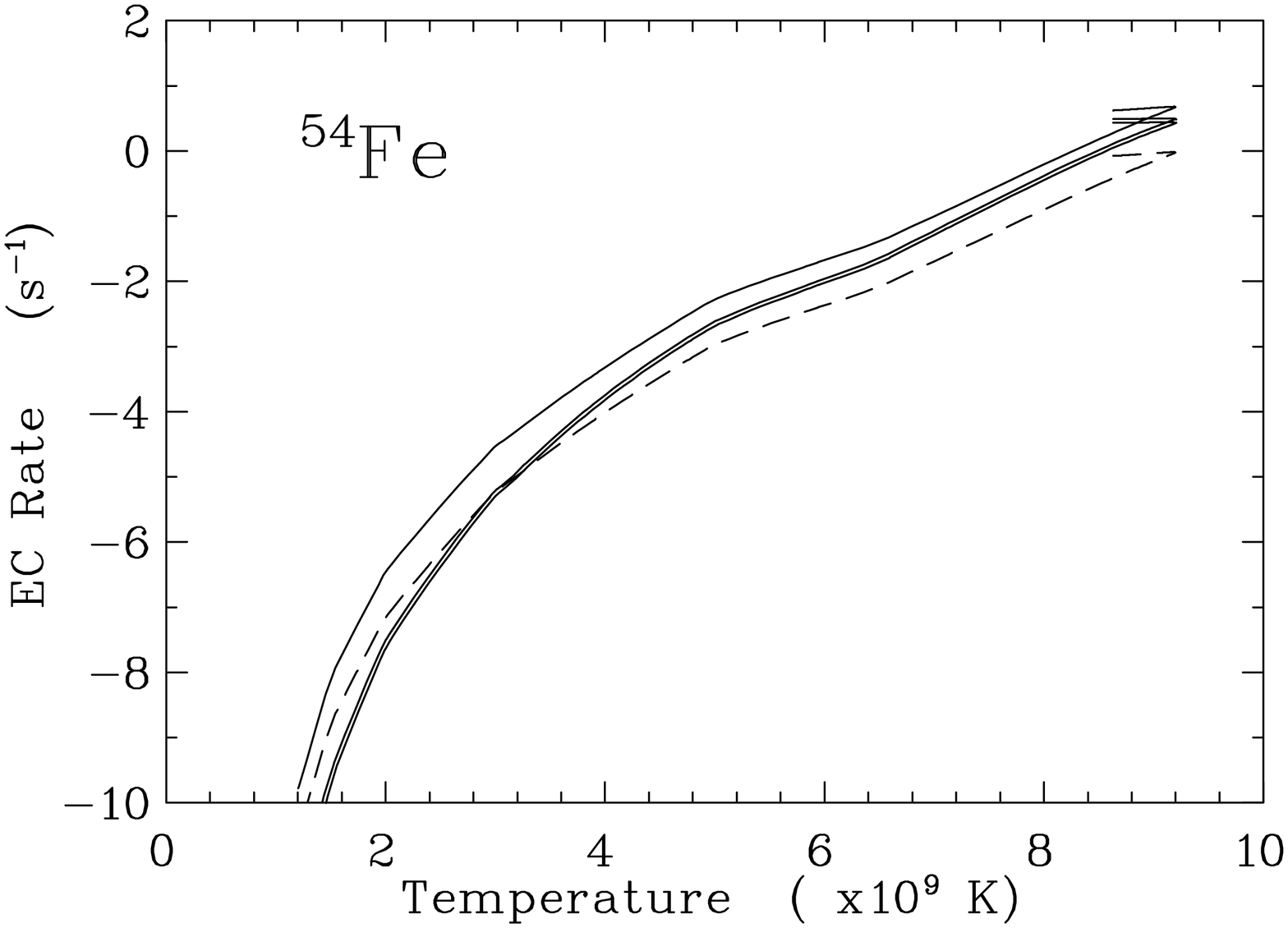}
    \includegraphics[width=4cm,angle=90]{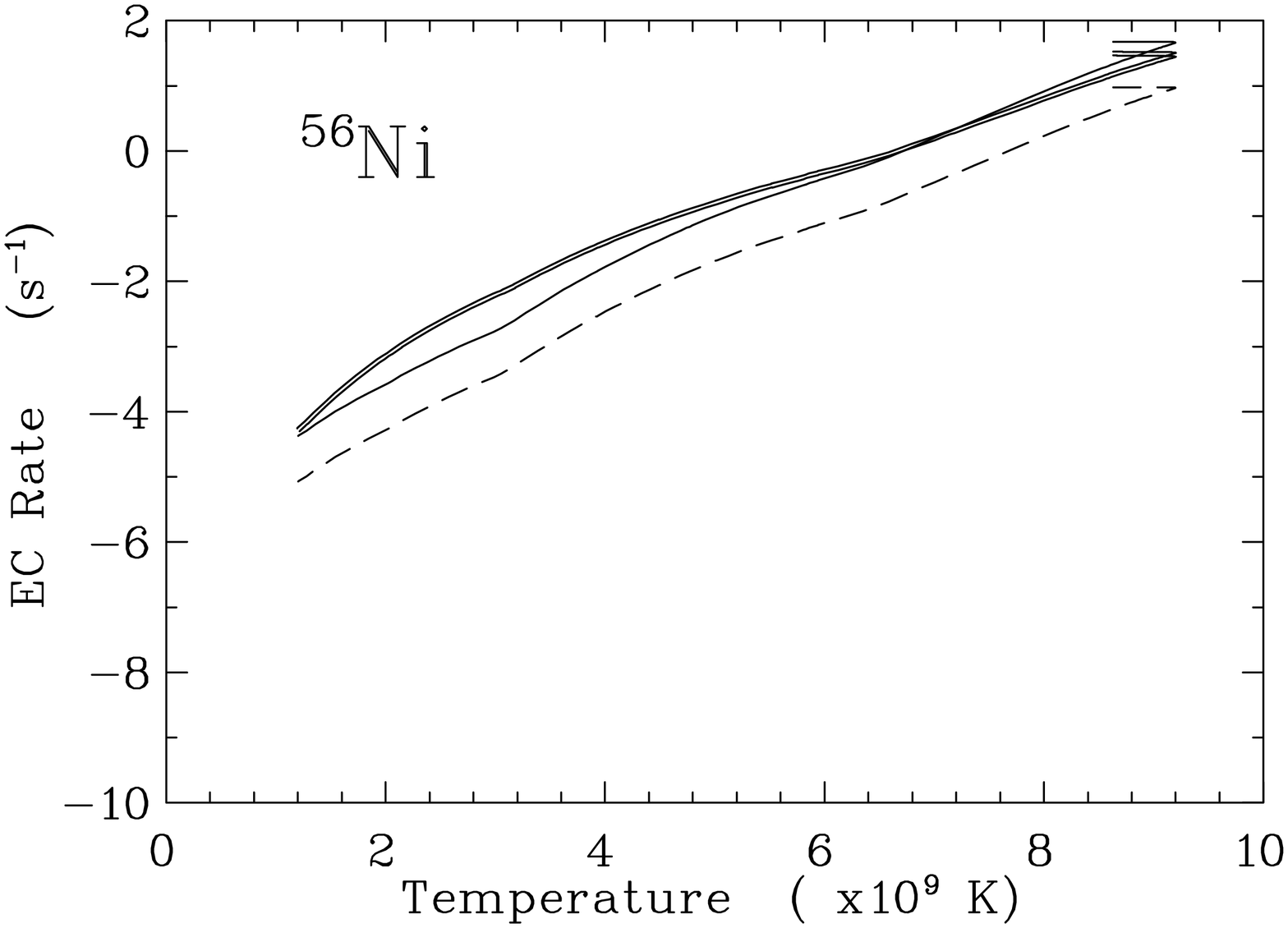}
    \includegraphics[width=4cm,angle=90]{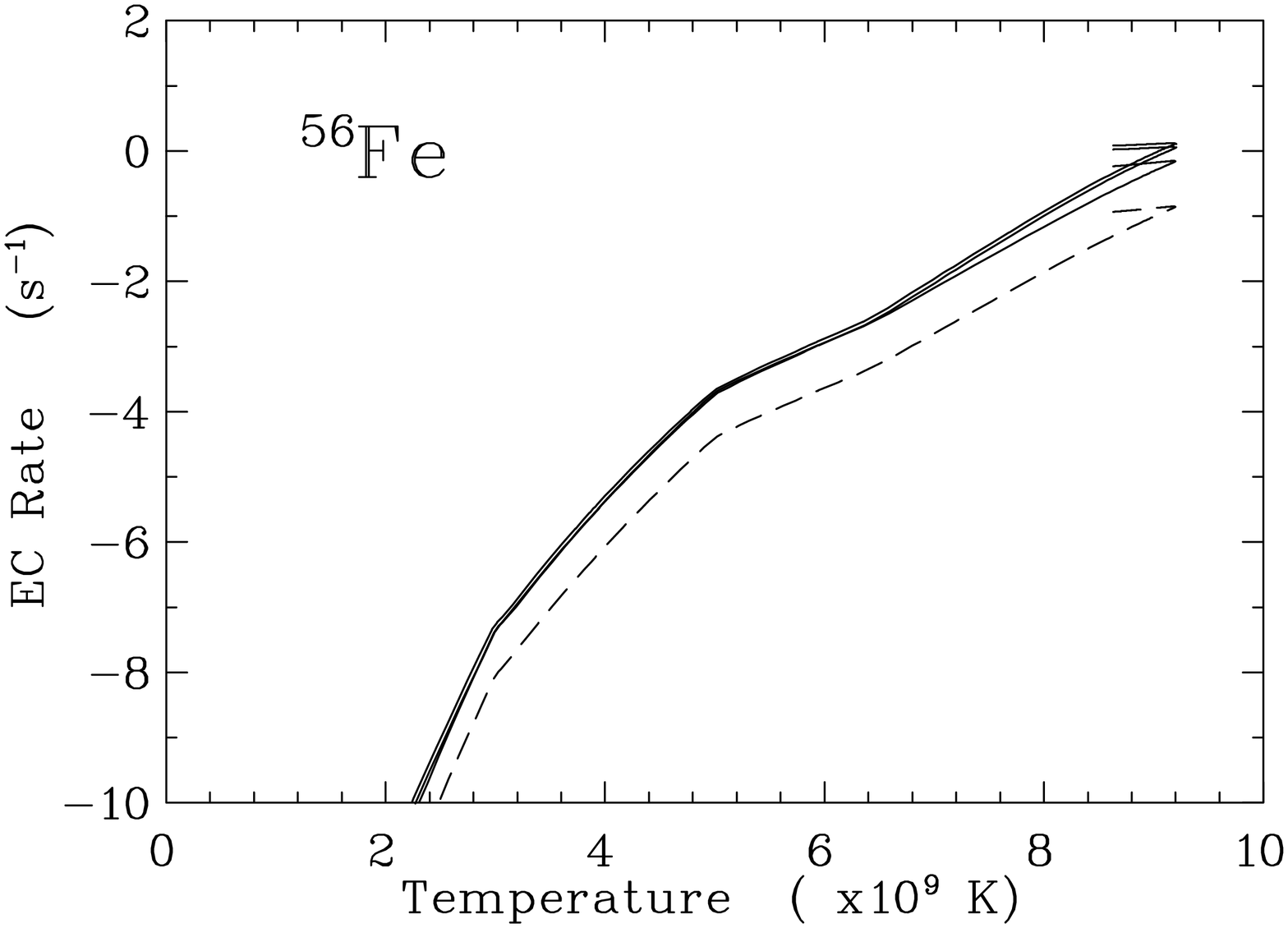}
    \includegraphics[width=4cm,angle=90]{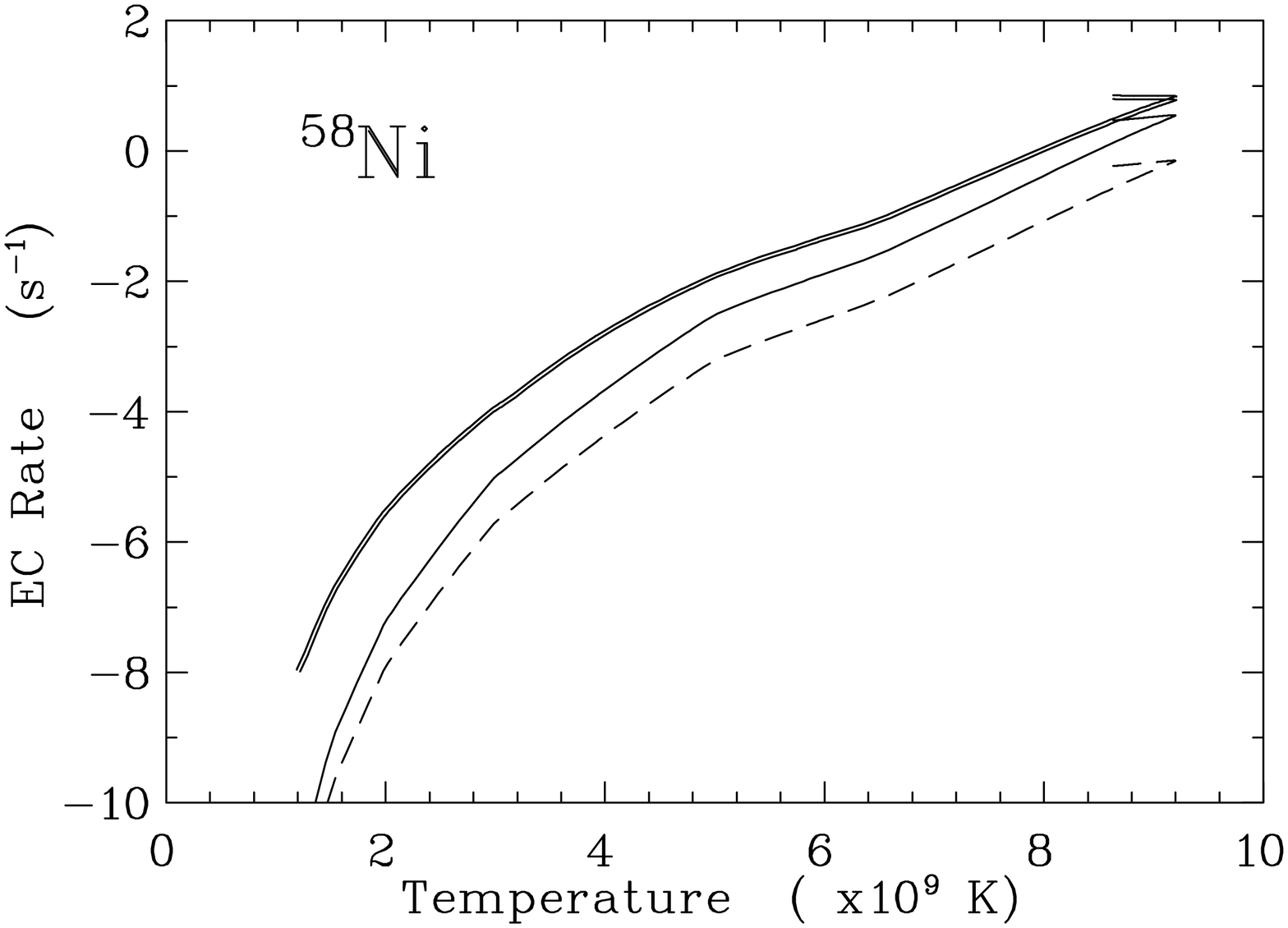}
    \includegraphics[width=4cm,angle=90]{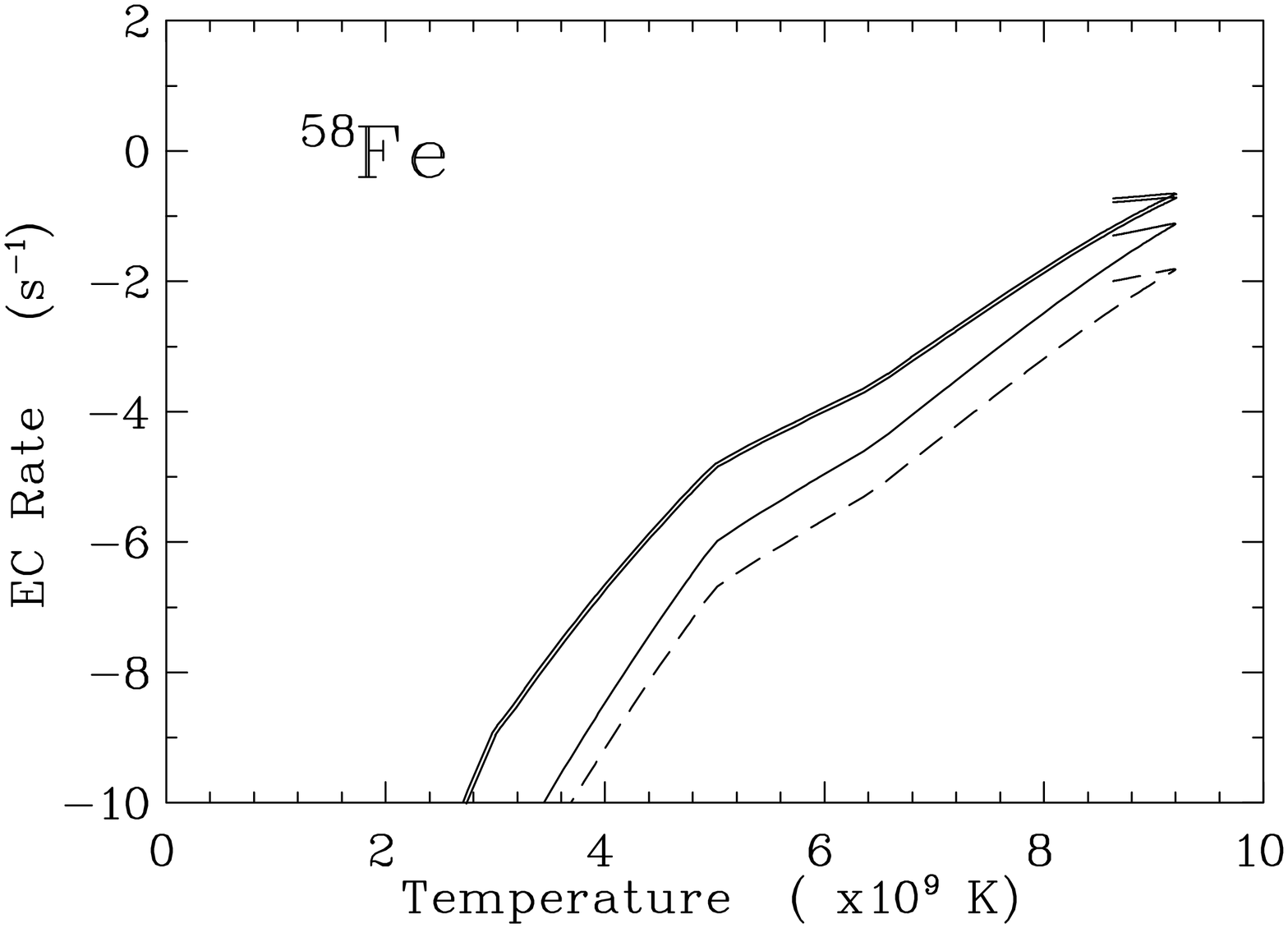}
    \includegraphics[width=4cm,angle=90]{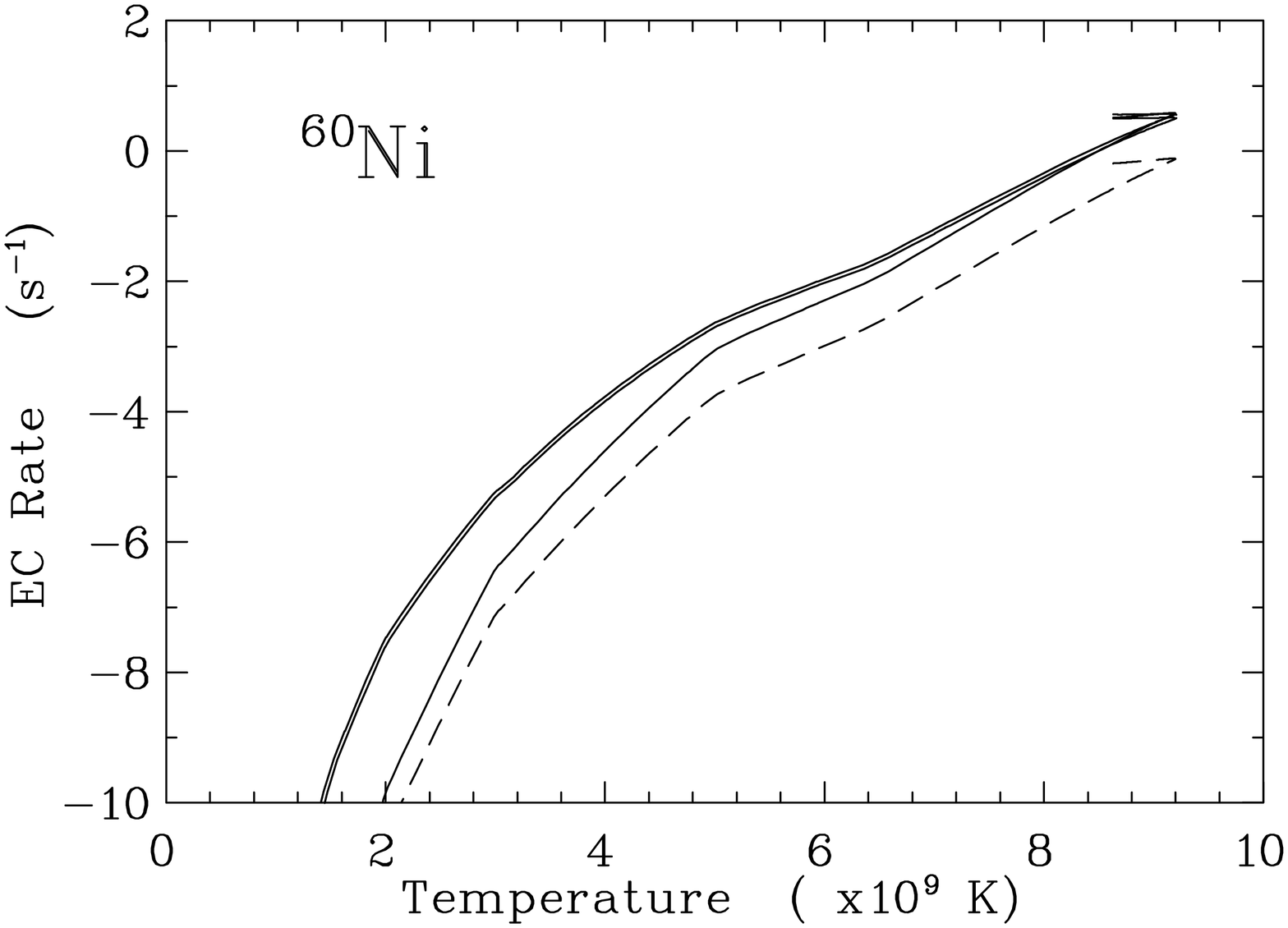}
    
    \caption{Electron capture rates [in s$^{-1}$] for even-even nuclei from 
    different sources 
    (FFN = Fuller et al.; Shell Model Monte Carlo SMMC = Dean \etal\ 1998;
    SMFA = estimates based on large-scale shell model diagonalization studies
    by Martinez \etal\ 1999). The rates are displayed as a function of 
    temperature only, but reflect the temporal variation of density and 
    temperature $\rho(t)$ and $T(t)$ in the trajectory of the innermost zone 
    of model WS15.
    \label{eecp}}
\end{center}
\end{figure}

\begin{figure}[htp]
    \includegraphics[width=5cm,angle=90]{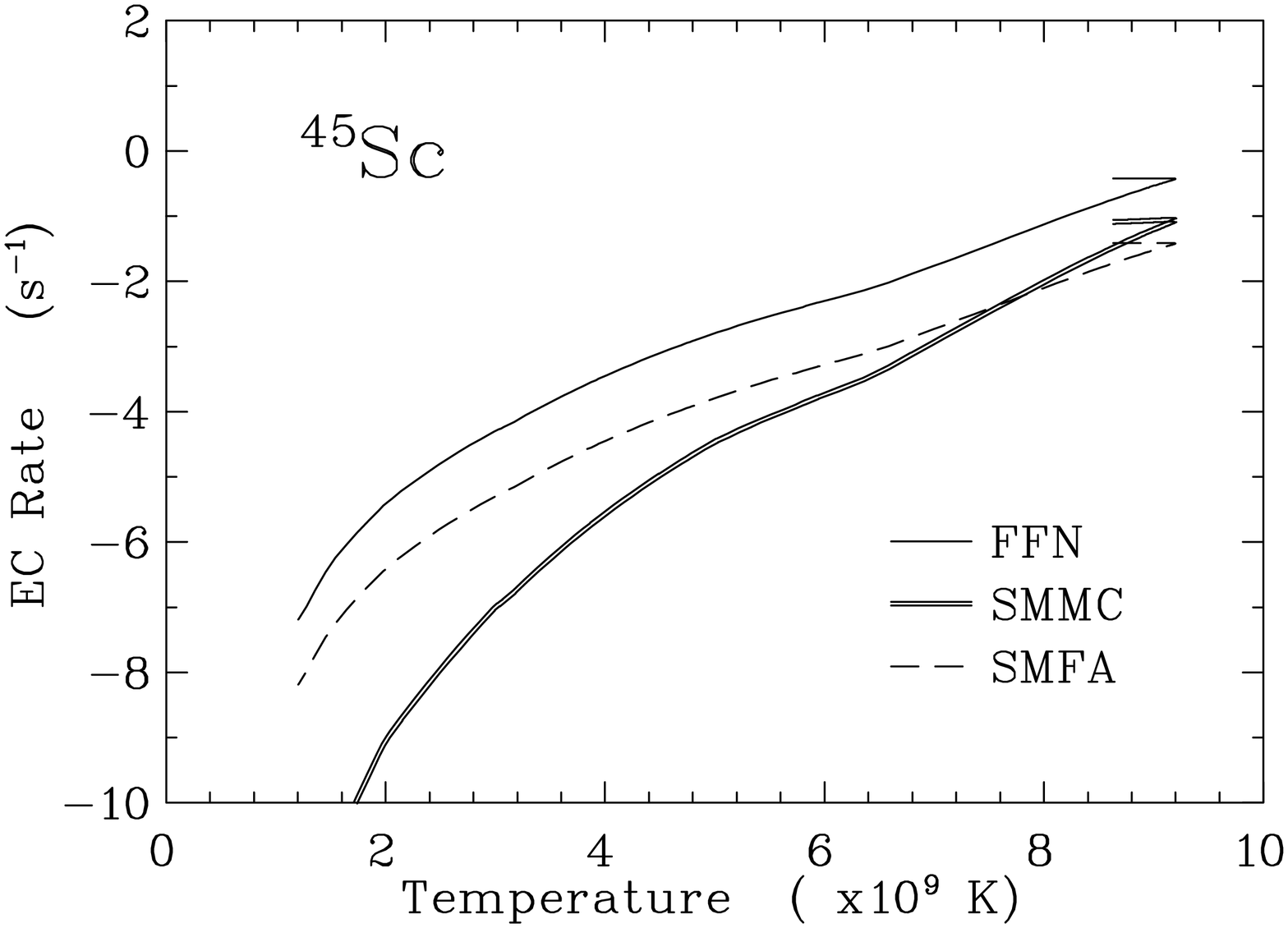}
    \includegraphics[width=5cm,angle=90]{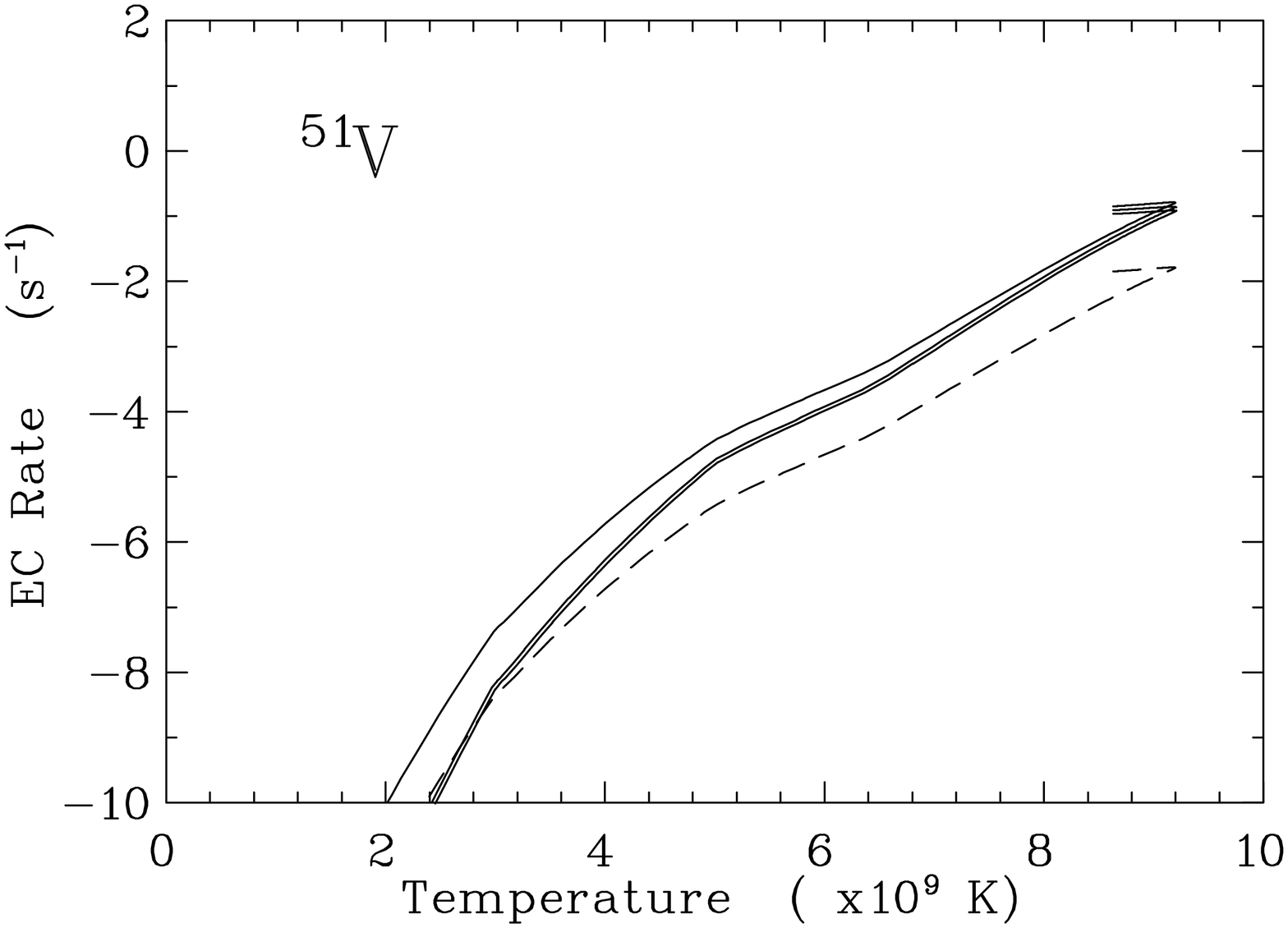}
    \includegraphics[width=5cm,angle=90]{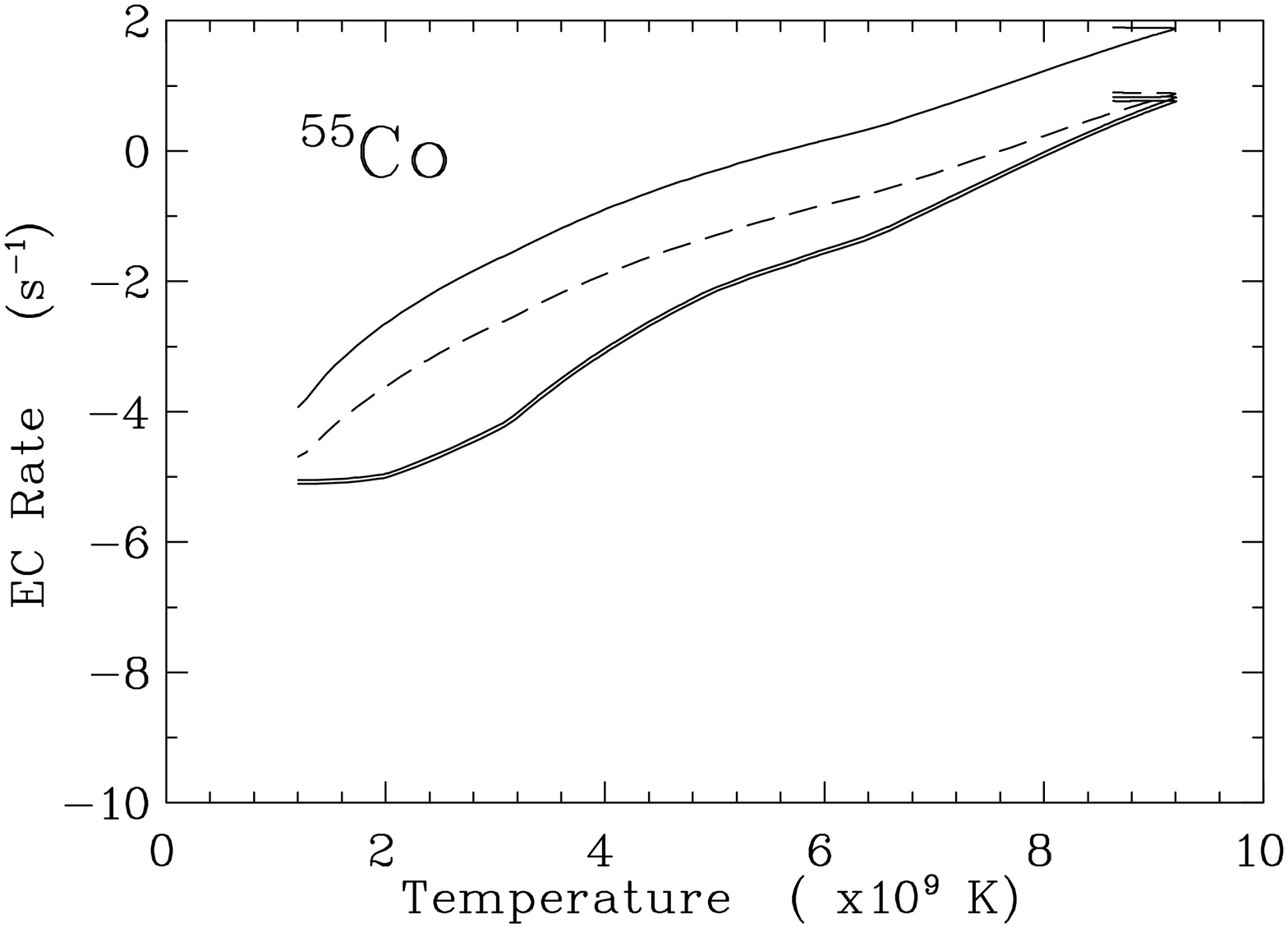}
    \includegraphics[width=5cm,angle=90]{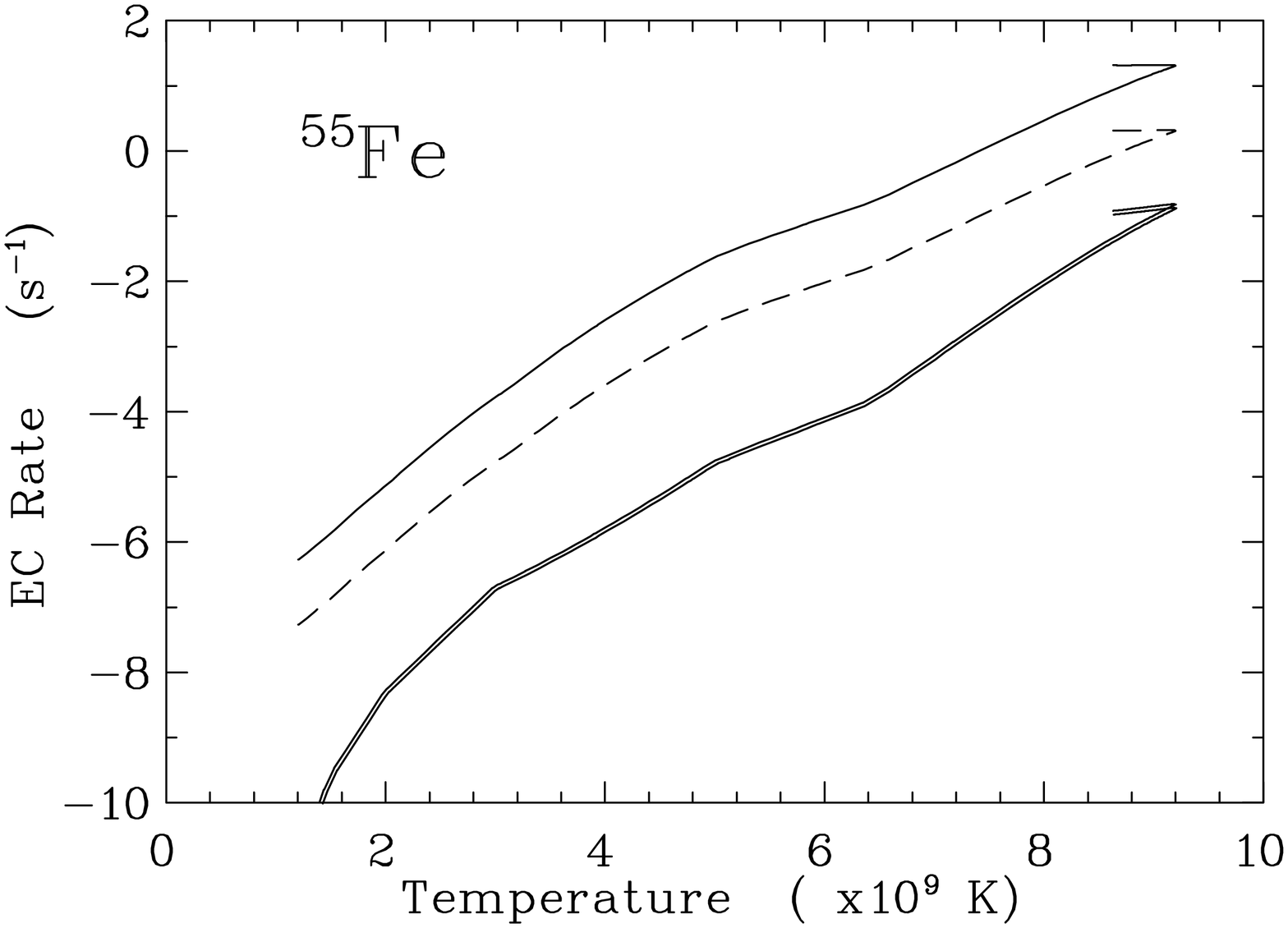}
    \includegraphics[width=5cm,angle=90]{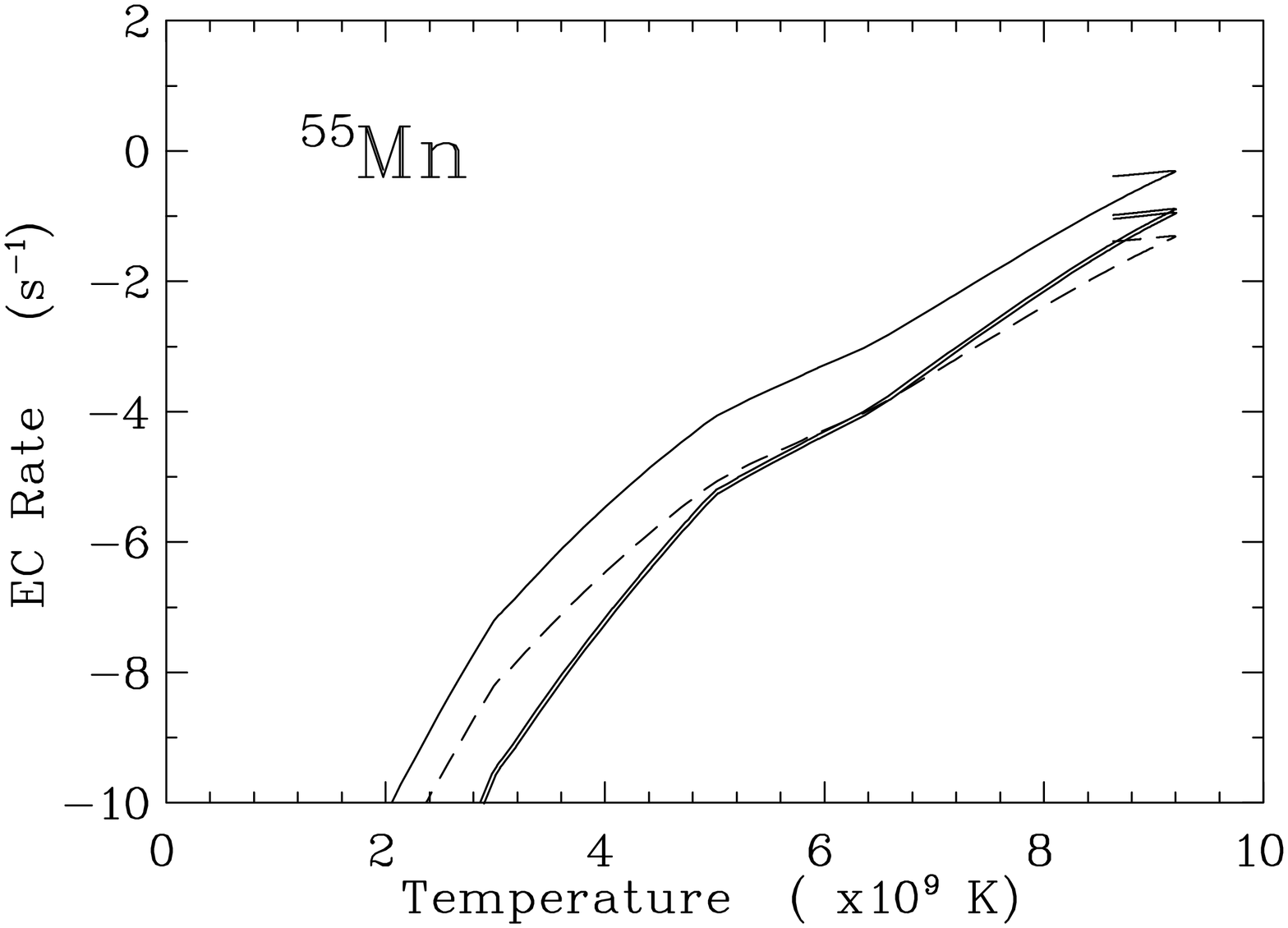}
    \includegraphics[width=5cm,angle=90]{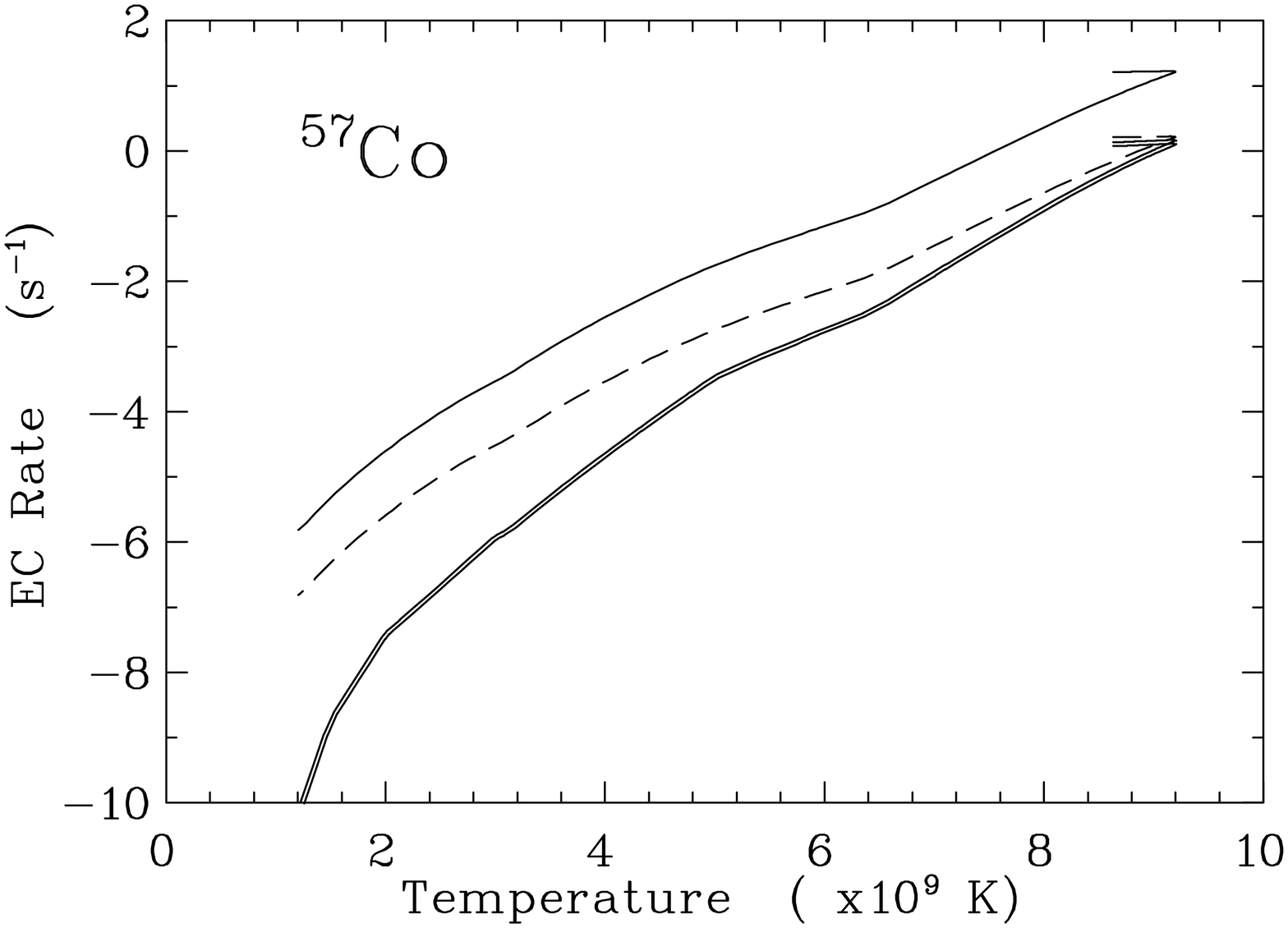}
    \includegraphics[width=5cm,angle=90]{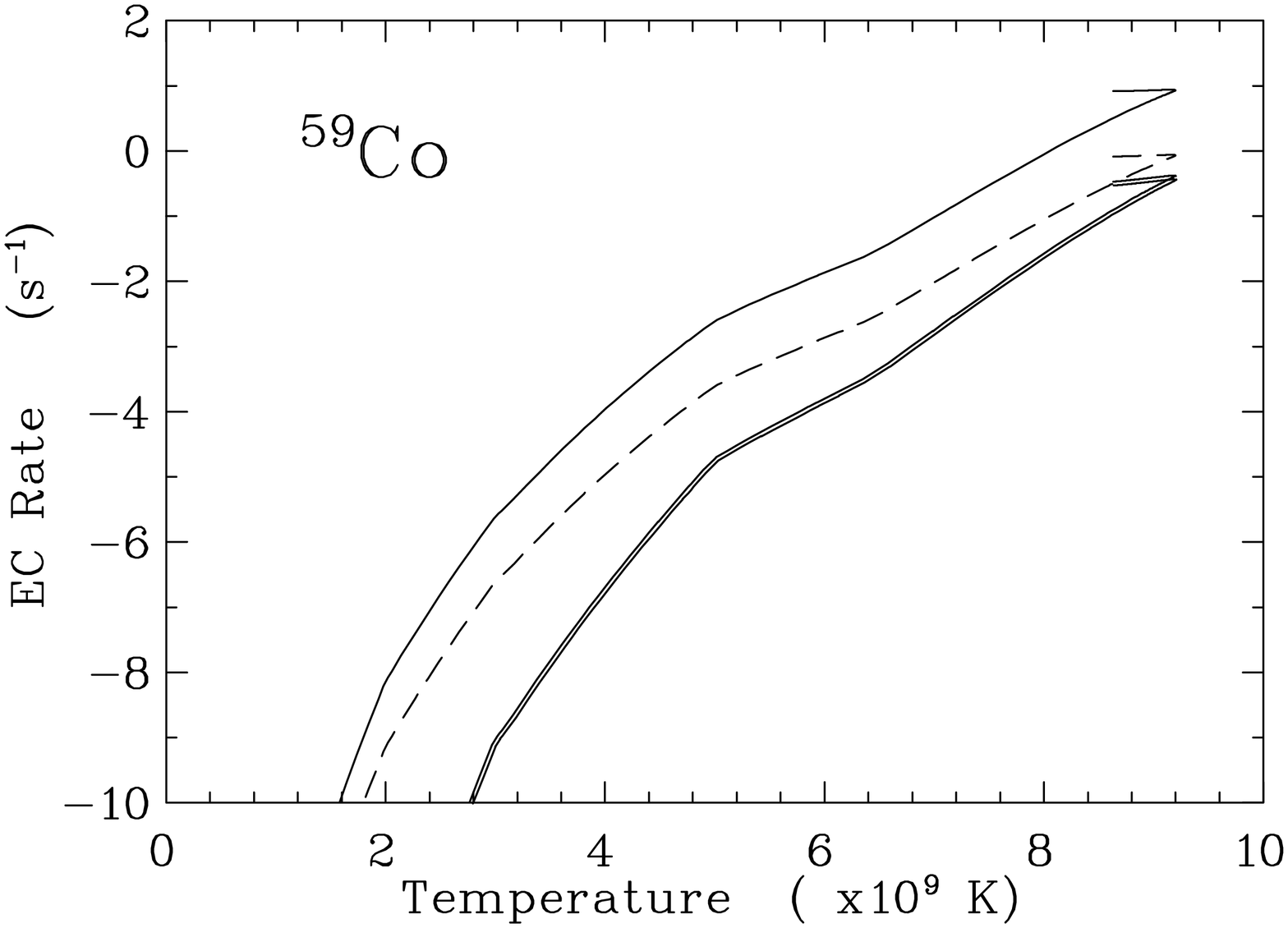}
    \caption{Same as Figure 1 for odd-A nuclei.
    \label{oocp}}
\end{figure}

In Figures \ref{eecp} and \ref{oocp} the different versions 
of electron capture rates are plotted as a function of temperature
for each nucleus for which SMMC electron capture rates are
available. As the rates are a function of temperature and density,
this includes implicitely the temporal evolution of the density 
up to the maximum temperatures attained, 
taken from the trajectory ($T(t),\rho(t))$ of the innermost zone of 
the SN Ia model WS15 discussed
in the following section. For the SMMC rates  the systematic
deviation from FFN are seen as mentioned earlier. For the even-even
parent nuclei (Figure \ref{eecp}) both rates are approximately the same, at 
least when approaching higher temperatures. For the odd-A nuclei 
(Figure \ref{oocp}) SMMC rates are definitely smaller. The SMFA electron 
capture rates are also smaller
than the FFN rates for both kinds of parent nuclei. When comparing SMMC
and SMFA we see, that
for odd-A nuclei the SMFA electron capture rates usally range
(with the exception of $^{51}$V) between FFN and SMMC. 
The fact that the SMFA rates are larger than the SMMC rates shows the
importance of the low-lying transitions unresolved in
SMMC approach as discussed above.
For the even-even nuclei the SMFA rates are on average
smaller than the SMMC rates. 
One reason for this change is the improved version of the residual interaction
employed in \citet{Martinez99} in comparison to \citet{dean98}, as discussed
above. We do 
not show a comparison for odd-odd nuclei, because these rates are not available
from the SMMC studies. But we refer to the discussion in the previous
paragraph that there one expects the largest deviations.

\section{SNe Ia Explosion Calculations}
As shown in \citet{thinomyok86} and reanalyzed in great detail in our recent 
paper \citep{iwamoto99}, electron capture is active on free protons and 
Fe-group nuclei during
the early burning stage of a thermonuclear SN Ia explosion, when the burning
front passes through the central region.
Thus, electron capture neutronizes matter and
reduces $Y_e$ from its original value close to 0.5 (0.4989 if the
abundances of nuclei other than $^{12}$C and $^{16}$O are due to a solar 
metallicity). This leads to the production of nuclei in the range between 
$N$=$Z$ and stability ($N$$>$$Z$). Only in exceptional cases also nuclei are
produced which are more neutron-rich than stable species. This is also the
reason why, opposite to conditions in core collapse supernovae 
\citep{Martinez99}, $\beta^-$
decay does not play a prominent role. Thus, the most
neutron-rich nuclei encountered already before $\beta^+$-decay of unstable
isotopes include species such as $^{48}$Ca, $^{50}$Ti, $^{54}$Cr, and 
$^{58}$Fe. Less neutron-rich (but also stable) nuclei like $^{54}$Fe and 
$^{58}$Ni are produced for more moderate $Y_e$ values in the 
central zones of the SN Ia models.  In \citet{iwamoto99} we made use of 
the FFN electron capture rates to predict
the nucleosynthesis yields of Chandrasekhar mass models of SNe Ia.
Those models employed variations in $\rho_{ign}$, $v_{def}$, and $\rho_{tr}$.
As the outer (lower density) layers, where the deflagration-detonation 
transition can occur, are not affected by electron capture, we can
neglect the third parameter $\rho_{tr}$ in the present discussion.
Here we show how the new electron capture rates 
affect the nucleosynthesis yields of the models WS15 and CS15. 
Those two models have the same burning front velocity, S15 denoting a
slow deflagration of 1.5\% of 
the local sound velocity, and differ in the ignition density at thermonuclear 
runaway, 1.7 x10$^{9}$ gcm$^{-3}$ 
(C) and 2.1 x10$^{9}$ gcm$^{-3}$ (W).

\citet{thinomyok86} showed that electron capture on free protons (which are
not affected by uncertainties of pf-shell nuclei) dominates for the 
thermodynamic conditions in explosive burning zones of SNe Ia. In the case of 
the FFN rates they amount to about 60\%. Thus, if electron capture rates of
Fe-group nuclei are reduced by a factor of 10, this affects only the
remaining 40\%, and the full effect of a change in electron capture rates
is a reduction from 100\% to 64\%. The difference in $Y_e$ between the
initial (almost symmetric) value of 0.4989 and the final value after explosive
burning and electron captures is therefore reduced with respect to the FFN 
calculations by about a factor 0.64. This point is the key to understanding 
results from calculations with different sets of electron capture rates.

In our previous studies we noted that 
the ignition density is a quantity which greatly influences 
the amount of electron capture in the central layers.
The higher ignition density of WS15 increases the Fermi energy and therefore
the electron capture rates, which leads to a smaller $Y_{e}$ when compared to 
CS15. Therefore, WS15 synthesizes more
neutron-rich nuclei than CS15. With the FFN rates used in \citet{iwamoto99} 
this led 
to a strong overproduction of $^{50}$Ti and $^{54}$Cr in comparison to the 
corresponding solar values for WS15. Consequently model CS15 seemed to be
preferrable in avoiding overabundances of the neutron-rich nuclei.
While $\rho_{ign}$ shifts the average $Y_e$-value of the central layers,
the burning front speed $v_{def}$ determines the gradient of $Y_e(r)$.
Information passes with sound speed to the outer layers.
Here a small $v_{def}$ permits a longer time for expansion to lower densities 
before the
arrival of the burning front. This reduces the effect of electron capture
in the outer layers of the central core and steepens the $Y_e$ gradient.
Thus, while $\rho_{ign}$ is mostly responsible for the minimum $Y_e$-values
which are attained in the central layers, $v_{def}$ controls the amount of 
matter with intermediate $Y_e$-values (like e.g. $^{54}$Fe and $^{58}$Ni)
by determining the $Y_e$-gradient. We will analyze here
how variations in electron capture rates influence the conclusions
drawn earlier for $\rho_{ign}$ and $v_{def}$.

The rates employed in our calculations were as follows:
 (i) FFN rates as a
benchmark for the further comparison (these rates were taken from
\cite{FFNrates} and hence do not incorporate any quenching of the GT
strength); (ii)
inclusion of the electron captures rates calculated
within the SMMC method by replacing the corresponding FFN rates. 
SMMC rates were used for the parent nuclei $^{45}$Sc,$^{48,50}$Ti,
$^{51}$V,$^{50,52}$Cr,$^{55}$Mn,$^{54-56,58}$Fe, and $^{55,57,59}$Co,
$^{56,58,60}$Ni. For nuclei not mentioned above (where no SMMC 
calculations were available), the rates were taken from FFN;
(iii) to simulate potential modification of the rates not provided by the
SMMC method, we also multiplied the FFN electron capture and beta-decay rates
within the Fe-group nuclei by these factors, 
derived from comparison between FFN and shell model rates (see section
2). These modified SN Ia models are labeled with SMFA;
(iv) A further option is to treat even-even (ee), odd-A (oa), and odd-odd (oo)
nuclei in different
ways, in order to test the sensitivity of the models and the importance of
certain rates in particular nuclei.
Such calculations are denoted by SMFA with the corresponding extension
ee, oa, oo or by combinations, e.g. ee+oa.
With these modifications of the electron capture rates we recalculated 
the nucleosynthesis for
the SN Ia models WS15 and CS15. Resulting deviations from the FFN models 
of \citet{iwamoto99} are 
discussed in the following two subsections.

\subsection {Influence on  Abundances and $Y_{e}$-Patterns}
\begin{figure}[htp]
\begin{center}
    \includegraphics[width=7cm]{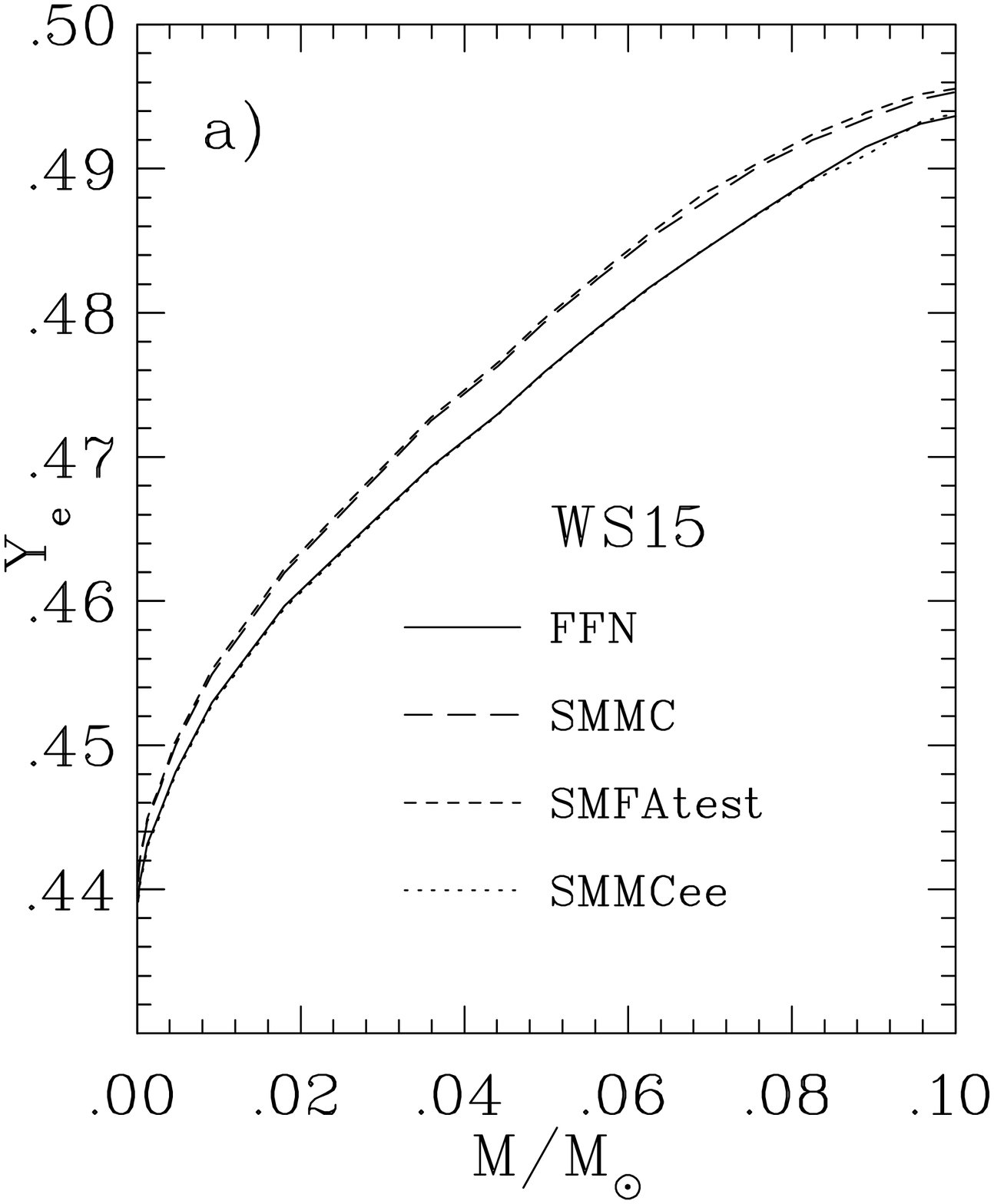}
    \includegraphics[width=7cm]{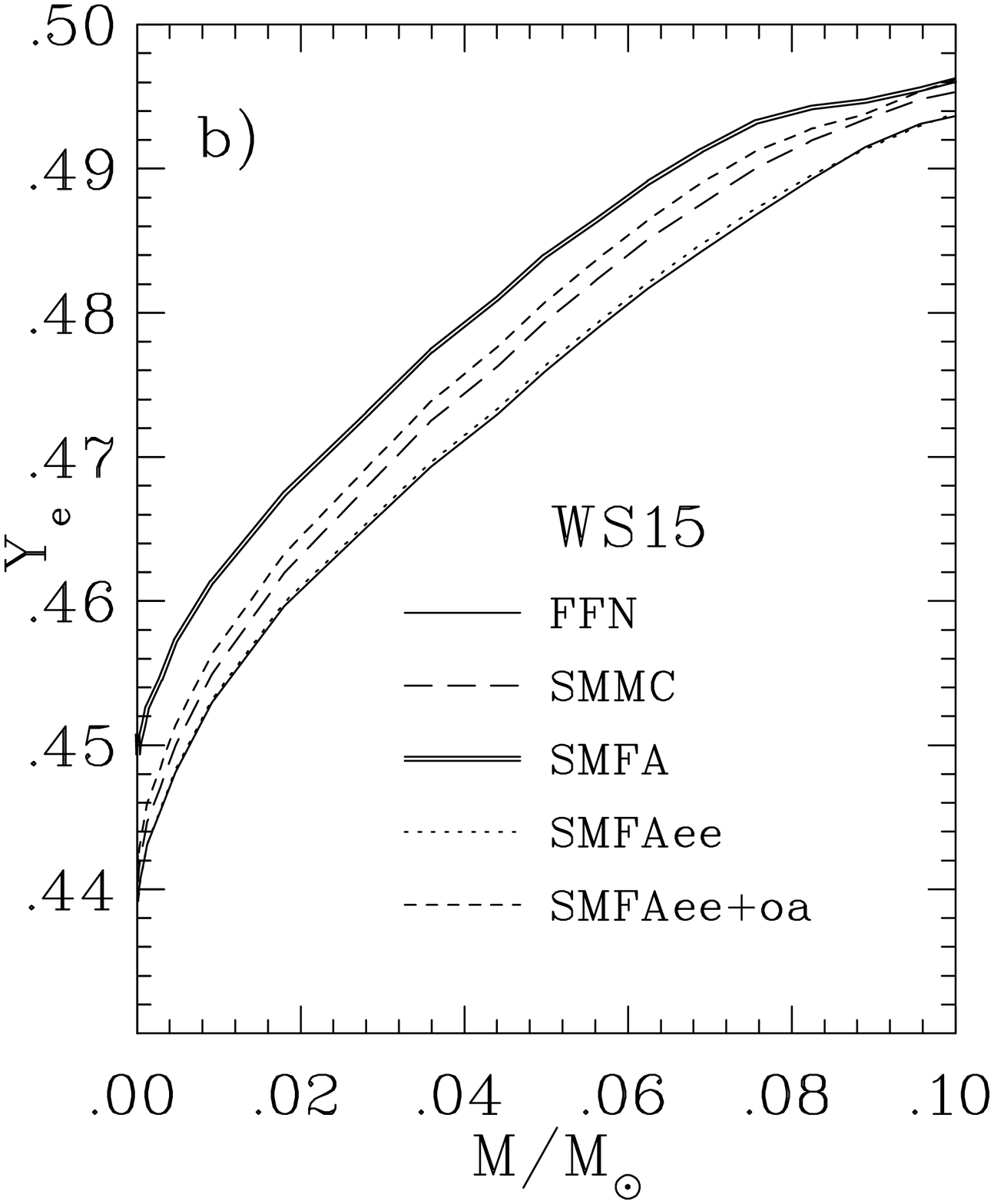}
    \includegraphics[width=7cm]{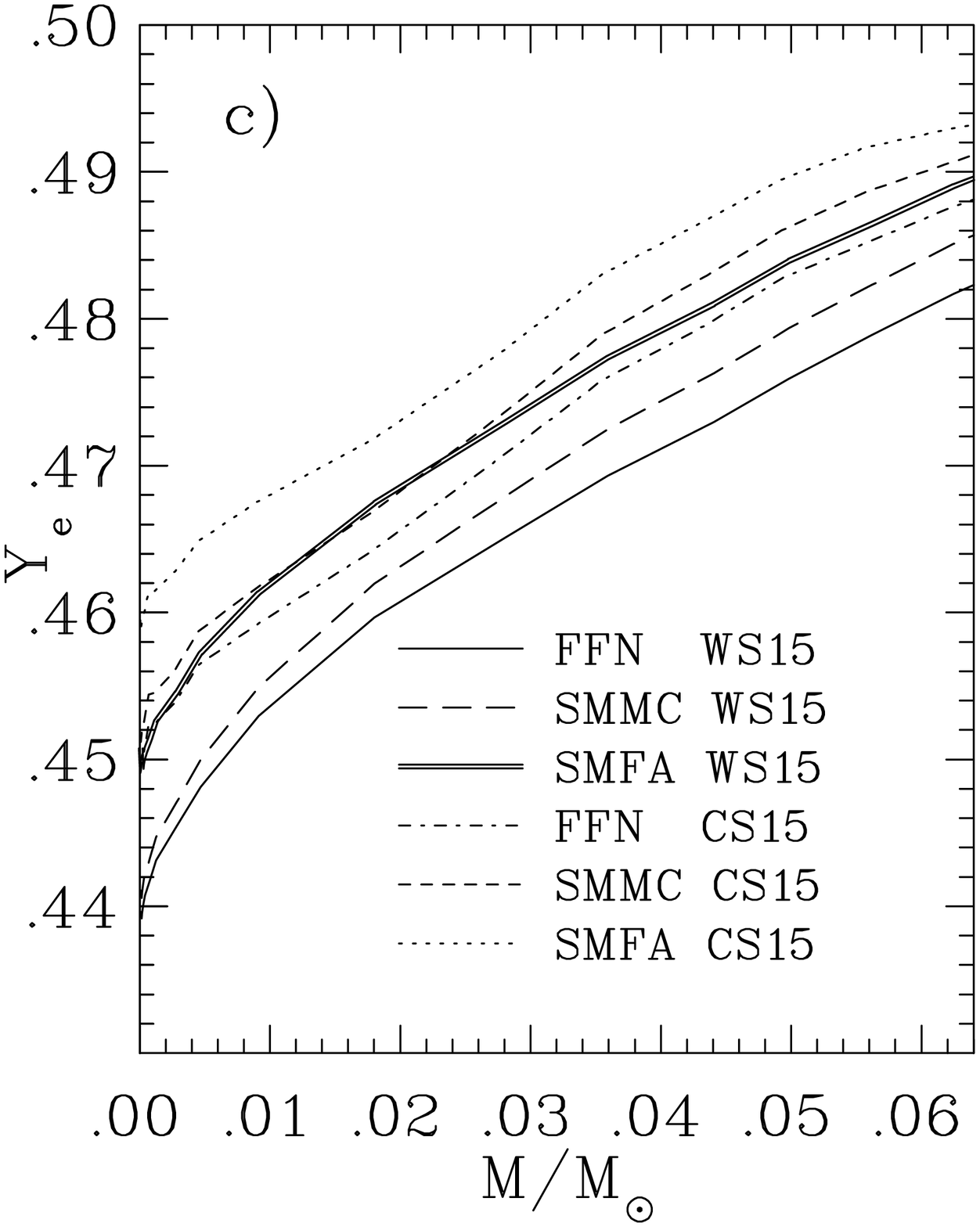}
    \includegraphics[width=7cm]{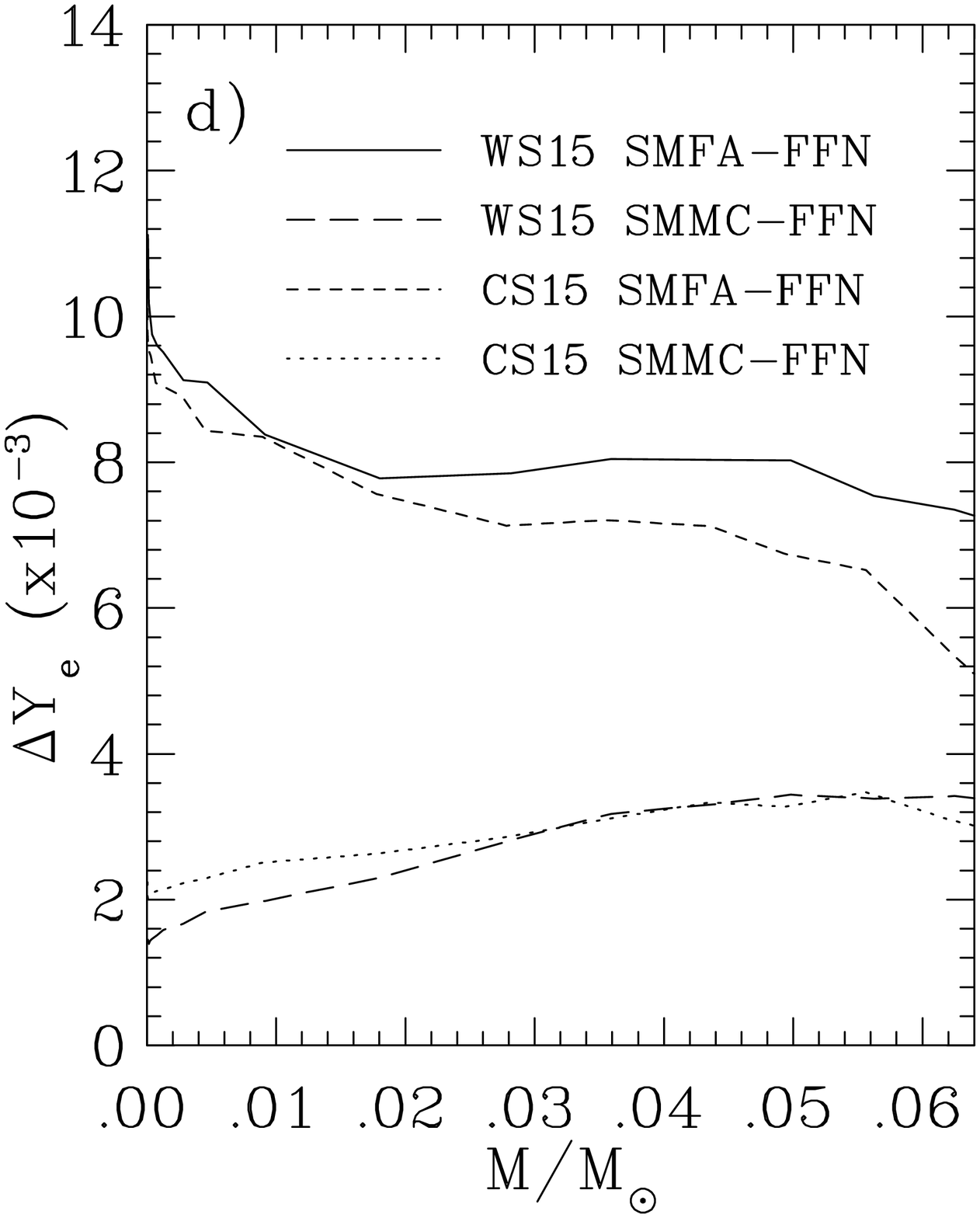}
    \caption{$Y_{e}$, the total proton to nucleon ratio and thus a measure of 
    electron captures on free protons and nuclei, after freeze-out of nuclear 
    reactions as a function of radial mass for different models and electron
    capture rates. Also the $Y_{e}$-difference  $\Delta Y_{e}$ between various
    cases is shown at the bottom right (d).  A detailed discussion of the 
    changes with different electron capture rate sets is given in the text. 
    Notice, however, that the changes for a given model (here WS15 and CS15)
    lead to almost parallel $Y_e$-curves in the intermediate $Y_e$-range
    responsible for the major abundances of $^{54}$Fe and $^{58}$Ni. This can 
    also be seen in the close to constant $\Delta Y_e$-curves in (d). Thus, 
    a change in electron capture rates does (to first order) not affect the 
    $Y_e$-gradient of a model. \label{yemr}}
\end{center}
\end{figure}

In general, the updated electron capture rates are smaller than  
the FFN rates. Thus the central region of the 
exploding white dwarf experiences less electron captures and 
the SN Ia nucleosynthesis yields should be less neutron-rich.
Therefore, $Y_{e}^{SMMC}$ and $Y_{e}^{SMFA}$ should be larger than 
$Y_{e}^{FFN}$ in the central layers and the overabundances of the neutron-rich
nuclei $^{54}$Cr and $^{50}$Ti in WS15 should be reduced.
The central $Y_{e}$-value, $Y_{e,c}$, of different models and 
different electron capture rate sets are listed in the Table \ref{tab1}. 
In the case of SMMC the final $Y_{e,c}$ value of the model WS15 increased 
from 0.440 (FFN) to 0.441. For the model CS15, which has a smaller 
ignition density leading to a higher $Y_{e,c}$ than WS15, the final 
$Y_{e,c}$-value changed from 0.449 (FFN) to 0.451 
(see Figure \ref{yemr}a and c).
On average the $Y_{e}^{SMMC}$-values for the central layers are about 0.002 
larger than $Y_{e}^{FFN}$ for both models (Figure \ref{yemr}d). 

\begin{table}[htp]
\caption{Neutron-rich Nucleosynthesis of specific nuclei with mass
fraction $X_i$; $X_i/X(^{56}{\rm Fe})/(X_i/X(^{56}{\rm Fe}))_\odot$}
\label{tab1}
\begin{center}
\begin{tabular}{lrrrrrrrr}
\hline
Model & $Y_{\rm e,c}$
& $^{50}$Ti & $^{54}$Cr & $^{58}$Fe & $^{64}$Ni & $^{62}$Ni & 
$^{54}$Fe & $^{58}$Ni \\
\noalign{\hrule}
WS15  FFN       &   0.4396 & 5.3  &  9.8  & 2.6  & 1.1   &  2.5 &  1.2 &   2.7 \\   
WS15  SMMC      &   0.4411 & 3.8  &  7.7  & 2.1  & 0.7   &  2.4 &  1.2 &   2.7 \\   
WS15  SMFAee+oa &   0.4424 & 2.9  &  6.4  & 1.8  & 0.5   &  2.3 &  1.2 &   2.6 \\   
WS15  SMFA      &   0.4507 & 0.6  &  2.3  & 1.0  & 0.01  &  2.0 &  1.2 &   2.6 \\   
CS15  FFN       &   0.4491 & 0.5  &  2.1  & 0.9  & 0.01  &  1.9 &  1.3 &   2.7 \\  
CS15  SMMC      &   0.4513 & 0.2  &  0.9  & 0.5  & 0.004 &  1.8 &  1.2 &   2.7 \\ 
CS15  SMFA      &   0.4594 & 0.008&  0.008& 0.007& 0.0002&  1.6 &  1.2 &   2.6 \\  
\hline
\end{tabular}
\end{center}
\end{table}

Bearing in mind that SMMC electron capture rates are 
available for certain even-even and odd-A nuclei only,
and that for the latter it is difficult to resolve the transitions at
low-lying excitations in the SMMC approach,
it is important to test which type of nuclei are mostly responsible for the 
resulting $Y_{e}$-shift.  One of the tests involves the application of SMMC 
rates only for even-even nuclei,
while using FFN rates for all other nuclei. 
This is denoted by the label SMMCee. 
It is evident from the $Y_{e}$-curve in Figure \ref{yemr}a 
that this case is almost identical with FFN, implying that the major cause for 
the difference between FFN and SMMC is due to capture on odd-A nuclei. 
This underlines that electron capture on even-even nuclei seems unimportant
despite their large abundances in nuclear statistical equilibrium. The
reason is that rates for even-even nuclei (with which these abundances have to 
be multiplied) are very small due to large energy thresholds.

In order to investigate the size of the $Y_e$-changes resulting from 
uncertainties between SMMC and the
average factors (SMFA) deduced from large-scale shell model calculations,
we replaced the 17 SMMC electron capture rates by 
the corresponding scaled FFN rates (labeled SMFAtest). 
Figure \ref{yemr}a shows that both $Y_{e}$-curves are very similar.
Therefore, it seems that the SMFA and SMMC odd-A rates yield comparable
$Y_e$-results, indicating a similar behavior. This can be explained
by the fact that most of the electron captures occur at high 
temperatures where the low-lying GT strength (which differs in both
SMMC and SMFA approaches) is less important. As a result, for odd-A nuclei the 
use of SMFA factors or the Monte Carlo shell model
(SMMC) leads to similar results. (The same holds true for even-even rates,
as shown above).

When scaling {\it all} FFN rates that are used in the network as suggested
by the shell model diagonalization calculations (label SMFA),
i.e. not only those for which SMMC rates were available,
we see in Figure \ref{yemr}b, c, and d that $Y_{e}^{SMFA}$ is about 0.008 
larger than $Y_{e}^{FFN}$ for WS15 as well as CS15. The central $Y_{e}$-value 
increased
from 0.440 to 0.451 (WS15) and from 0.449 to 0.459 (CS15), as listed in Table
\ref{tab1}. 
The deviation from FFN is much larger than for SMMC.
This could have two possible causes.
(i) A more complete set of modified electron 
capture rates is used, 17 (SMMC) versus 79 (SMFA).
(ii) Odd-odd parent nuclei are missing in the SMMC calculations and could be
important. In order to
test which aspect plays the more important role,
we chose a subset where only even-even and odd-A nuclei were multiplied with
the average SMFA factors (labeled SMFAee+oa in Figure 3b). Thus, this
case ignores modifications for odd-odd nuclei, while the multiplcation
by SMFA factors for even-even and odd-A nuclei should differ little from
the use of SMMC rates as shown in the previous paragraph. Therefore, the 
comparison of SMFAee+oa with SMMC measures the impact of the
increased number of modified electron capture rates, and the comparison
to SMFA shows the influence of odd-odd nuclei.
The resulting $Y_{e}$-curve (Figure \ref{yemr}b) displays a small 
$Y_{e}$ shift between SMMC and SMFAee+oa, and a larger 
$Y_{e}$-shift between SMFAee+oa and SMFA.
Therefore, the inclusion of odd-odd nuclei has the largest influence on 
the $Y_{e}$ difference between SMFA and SMMC.

Thus, we have shown that the rate change for odd-A nuclei is mostly responsible
for the $Y_{e}$-shift between FFN and SMMC, and that the inclusion of odd-odd 
nuclei causes the largest part of the $Y_{e}$-shift between SMMC and SMFA. 
This makes clear that the changes in the electron capture rates for
odd-A and odd-odd nuclei are responsible
for the $Y_{e}$ difference between SMFA and FFN, while the contribution
of even-even nuclei is negligible, an assertion which was directly tested by
case SMFAee (Figure \ref{yemr}b). As odd-odd nuclei are difficult to treat 
within the shell model Monte Carlo approach, a further improvement would be the
direct use of large-scale shell model diagonaliztion calculations. In the
present paper we provide preliminary results by applying average factors
(SMFA) derived from detailed calculations of a few key nuclei \citep{Martinez99}

\begin{figure}[htp]
\begin{center}

    \includegraphics[height=.4\textheight,angle=90]{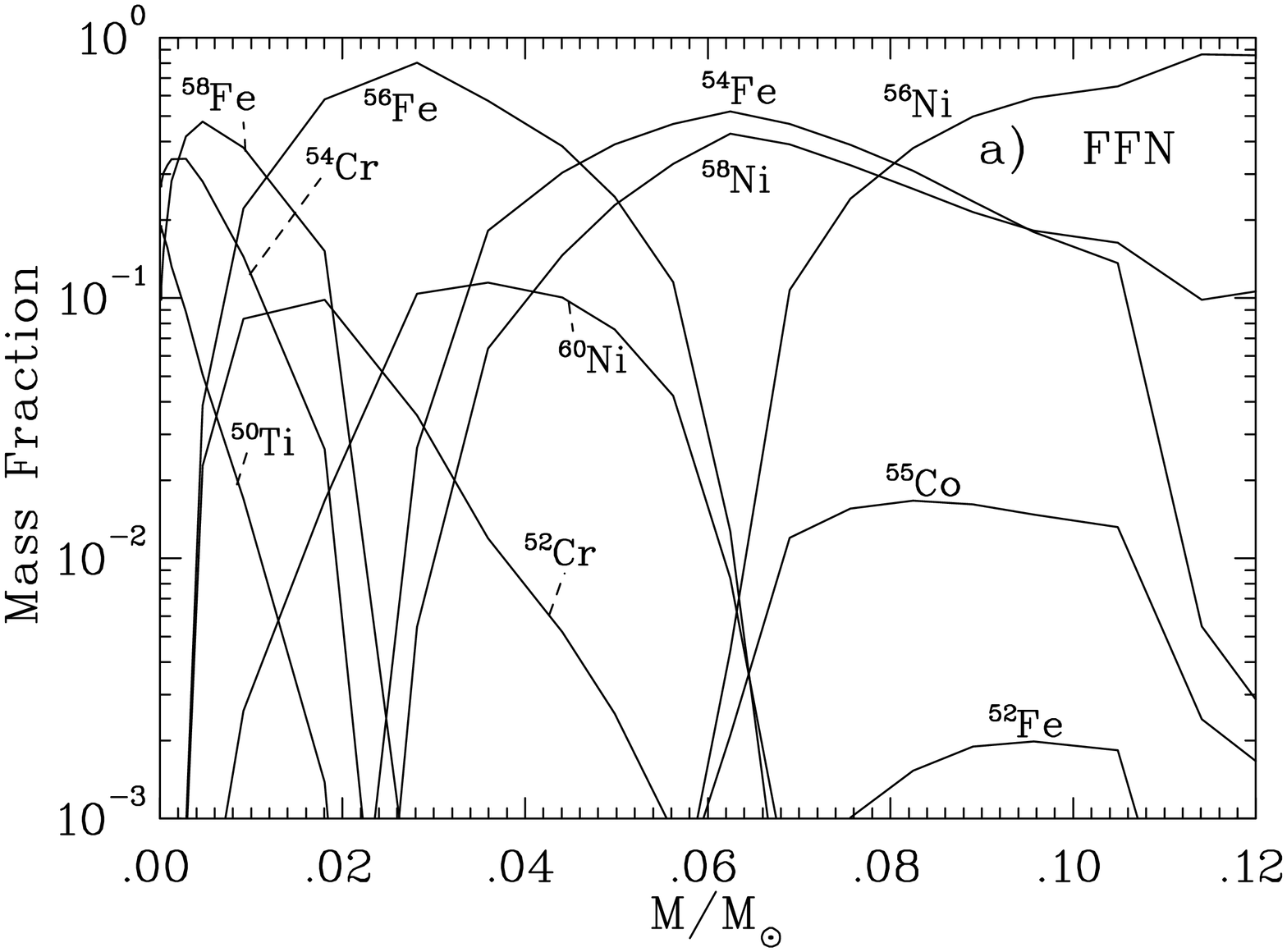}
    
    \includegraphics[height=.4\textheight,angle=90]{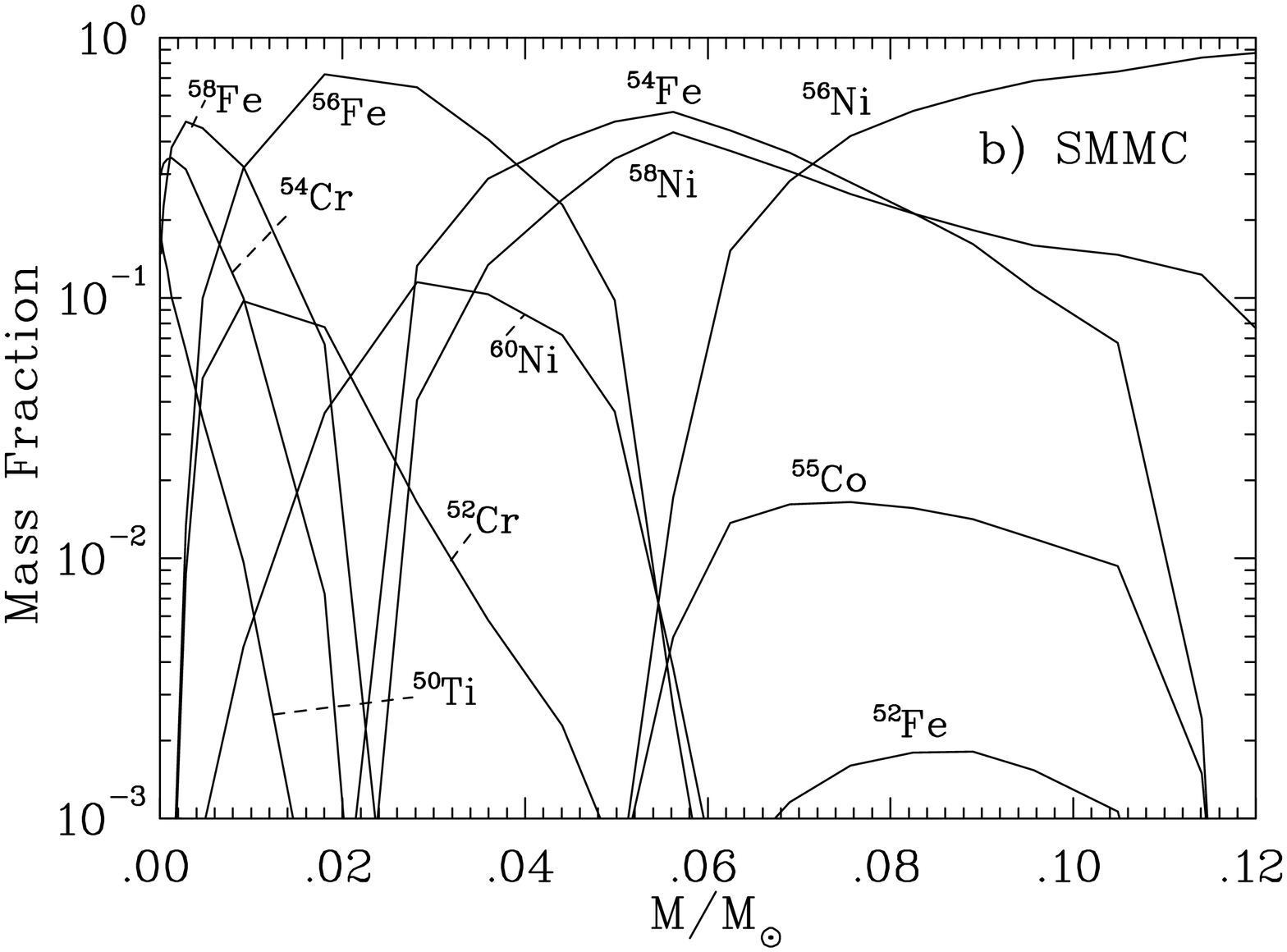}
    
    \includegraphics[height=.4\textheight,angle=90]{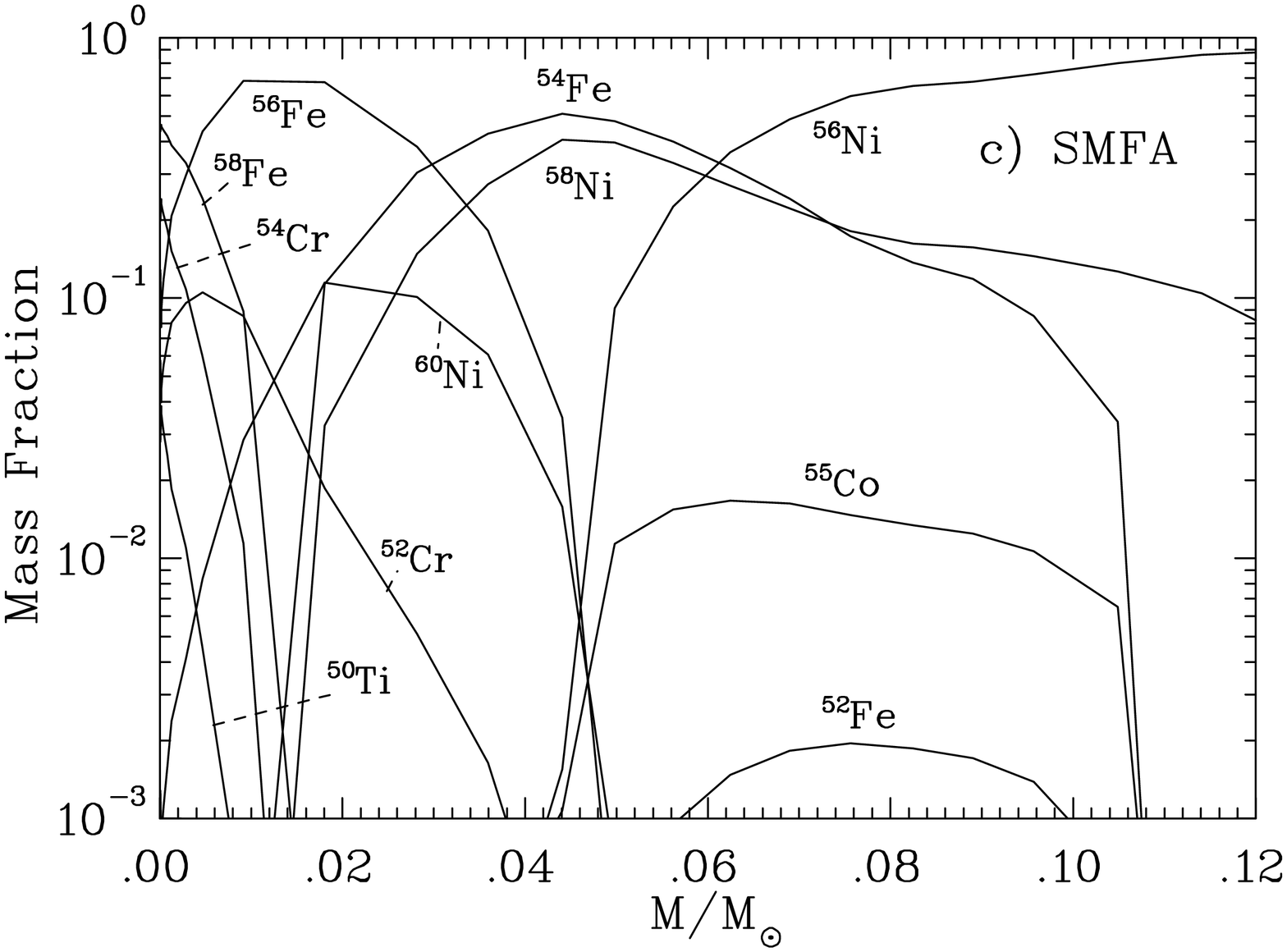}
    \caption{Central abundances in model WS15 for different electron capture 
    rates: (a) FFN, (b) Shell Model Monte Carlo (SMMC), (c) FFN multiplied 
    with average factor estimates from large-scale shell model diagonalization 
    studies (SMFA).
    \label{xi2}}
\end{center}
\end{figure}

To examine the impact of these changes in weak rates on individual species,
we show in Figure \ref{xi2} the radial distribution of a few key abundances 
for the three select cases FFN, SMMC, and SMFA. The abundance 
pattern is very similar, but each abundance curve is
shifted inwards in the sequence FFN, SMMC, and SMFA. This makes clear that
FFN reaches the smallest central $Y_e$'s, resulting in abundance peaks of
$^{50}$Ti, $^{54}$Cr, and $^{58}$Fe close to the center, while these
peaks are cut off for SMFA. Neglecting this very central behavior, one
can recognize, however, that the total amount of intermediate $Y_e$
nuclei like $^{54,56}$Fe and $^{58}$Ni is essentially unaffected
(see also Table \ref{tab1}). This is in accordance with the results of
Figure \ref{yemr}, where we see that the $Y_e$-gradient is almost idential
and the $\Delta Y_e$'s close to constant for models which apply different sets 
of electron capture rates. Note
that only the onset of the $Y_e$-reduction due to electron captures is 
shifted as a function of $M(r)$. This leaves
the same amount of mass composed of these intermediate $Y_e$ nuclei. The main
difference is in the central $Y_e$ values attained and as a result in the
amount of the most neutron-rich nuclei.

As the $Y_e$ gradient is determined by $v_{def}$ \citep[see][]{iwamoto99}
and apparently is not changed by different sets of electron capture rates,
we can conclude that the consequences for the permitted range of burning front
speeds remain the same. In \citet{iwamoto99} we determined this
range $v_{def}/v_s$ to be of the order 0.015-0.03. The central neutronization
is however dependent on $\rho_{ign}$ and, as shown here, on the
set of electron capture rates employed. Thus, we have to expect that
our previous conclusions for the $\rho_{ign}$-range might have to
be changed. We will discuss this further in the following subsection.

\subsection{Comparison to Solar Abundances}
The solar element abundances are a snapshot of the local galactic abundance 
distribution at the  formation of the solar system 4.5 x10$^{9}$ years ago.
The heavy elements in the Galaxy originate from the ejected matter of 
supernovae, with SN Ia. SNe Ia being responsible 
for 55\% or more of the Fe-group elements in the Galaxy
\citep[see the discussion in][]{iwamoto99}. Thus, the ejecta of SNe Ia can 
contain an overabundance in comparison with the solar values of no more than a 
factor of two among the Fe-group nuclei. This is the maximum for species which 
have no other production site than SNe Ia and would have to be reduced
accordingly if alternative sources contribute as well. This provides 
constraints for the nucleosynthesis results of SNe Ia models.
In \citet{iwamoto99}, where we used the FFN electron capture rates only, 
we concluded that CS15 was a better model compared to WS15 in terms of 
avoiding overproduction of neutron-rich nuclei. This would indicate that 
for the majority of white dwarfs undergoing a thermonuclear runaway and 
SNe Ia events, the central density should be lower than 
$\sim$2x10$^{9}$gcm$^{-3}$, though the exact constraint depends somewhat 
on the flame speed. 

Now, with the new sets of electron capture rates, the situation has changed. 
The new smaller rates reduce the production of neutron-rich nuclei. In 
Figures 5, 6, and 7 and in Table \ref{tab1} the ratios to solar abundances 
(normalized to $^{56}$Fe) are  displayed for WS15 and CS15 employing different 
types of electron capture rates. Here the results of the central slow 
deflagration studies have been merged with (fast deflagration) W7 compositions 
for the outer layers \citep{nomoto94,thinomyok86,thielemann97,iwamoto99}. In 
the outer layers the densities are sufficiently low that electron capture does
not modify the pre-explosion value of $Y_{e}$. 
($Y_e$ in these layers is only a witness of the initial metallicity of the 
white dwarf, manifesting itself in the amount of $^{22}$Ne, which resulted in 
H and He-burning from the initial CNO isotopes.)
Thus, the same model results for these outer layers can be added for 
different sets of electron capture rates. 

\begin{figure}[htp]
    \includegraphics[width=6cm,angle=90]{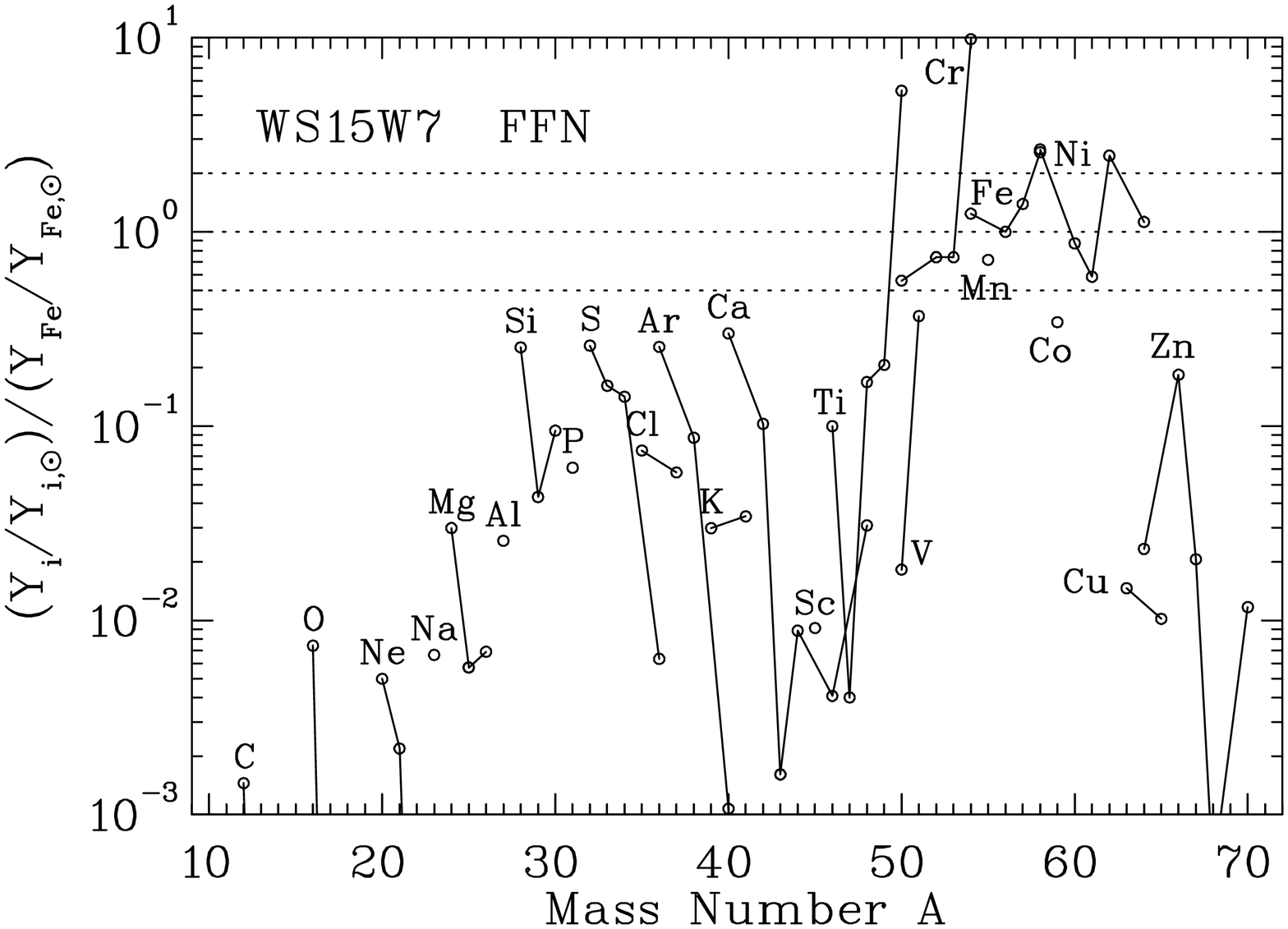}
    \includegraphics[width=6cm,angle=90]{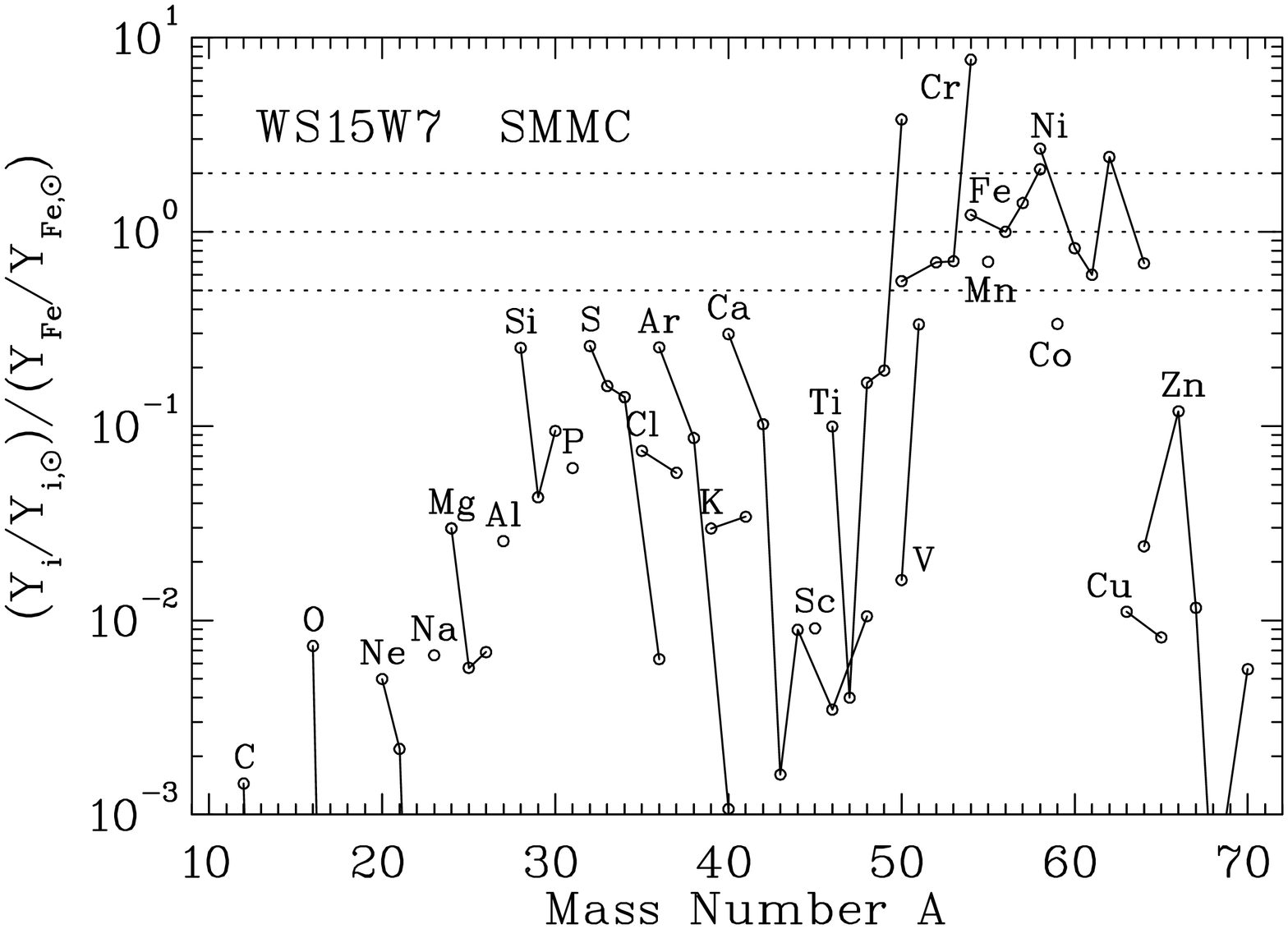}
    \caption{Ratio of abundances to solar predicted in model WS15 for different
    electron capture rate sets. Isotopes of one element are connected by lines. 
    The ordinate is normalized to $^{56}$Fe. Intermediate mass elements exist, 
    but are underproduced by a factor of 2-3 for SNe Ia models in comparison
    to Fe-group elements. The Fe-group does not show a composition close to
    solar. Especially $^{54}$Cr and $^{54}$Fe are strongly overproduced by 
    more than a factor of 3.  The change from FFN rates (top) to SMMC (bottom) 
    reduces the overproduction over solar to about 75\%. 
   \label{ysol1}}
\end{figure}

\begin{figure}[htp]
    \includegraphics[width=6cm,angle=90]{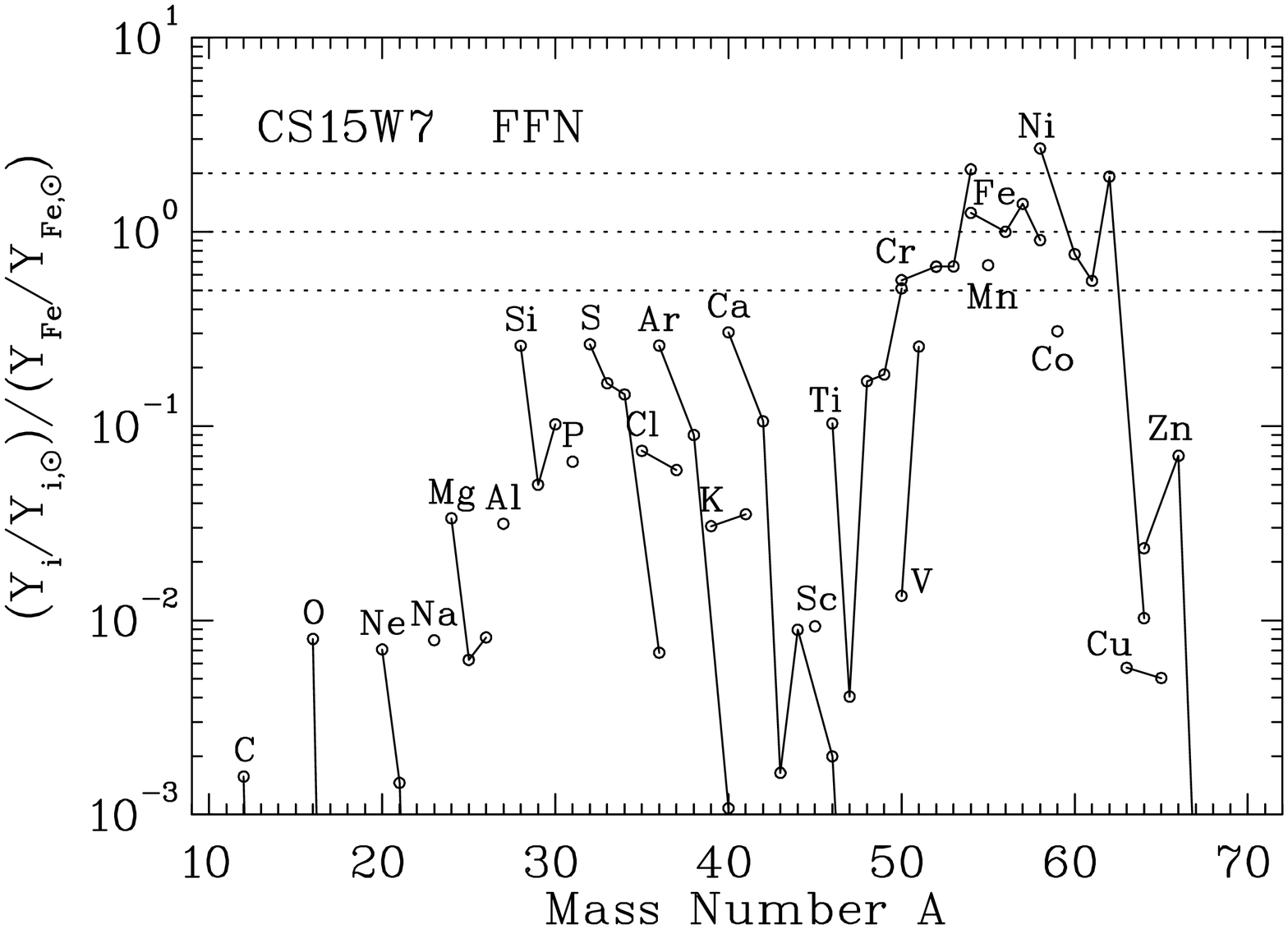}
    \includegraphics[width=6cm,angle=90]{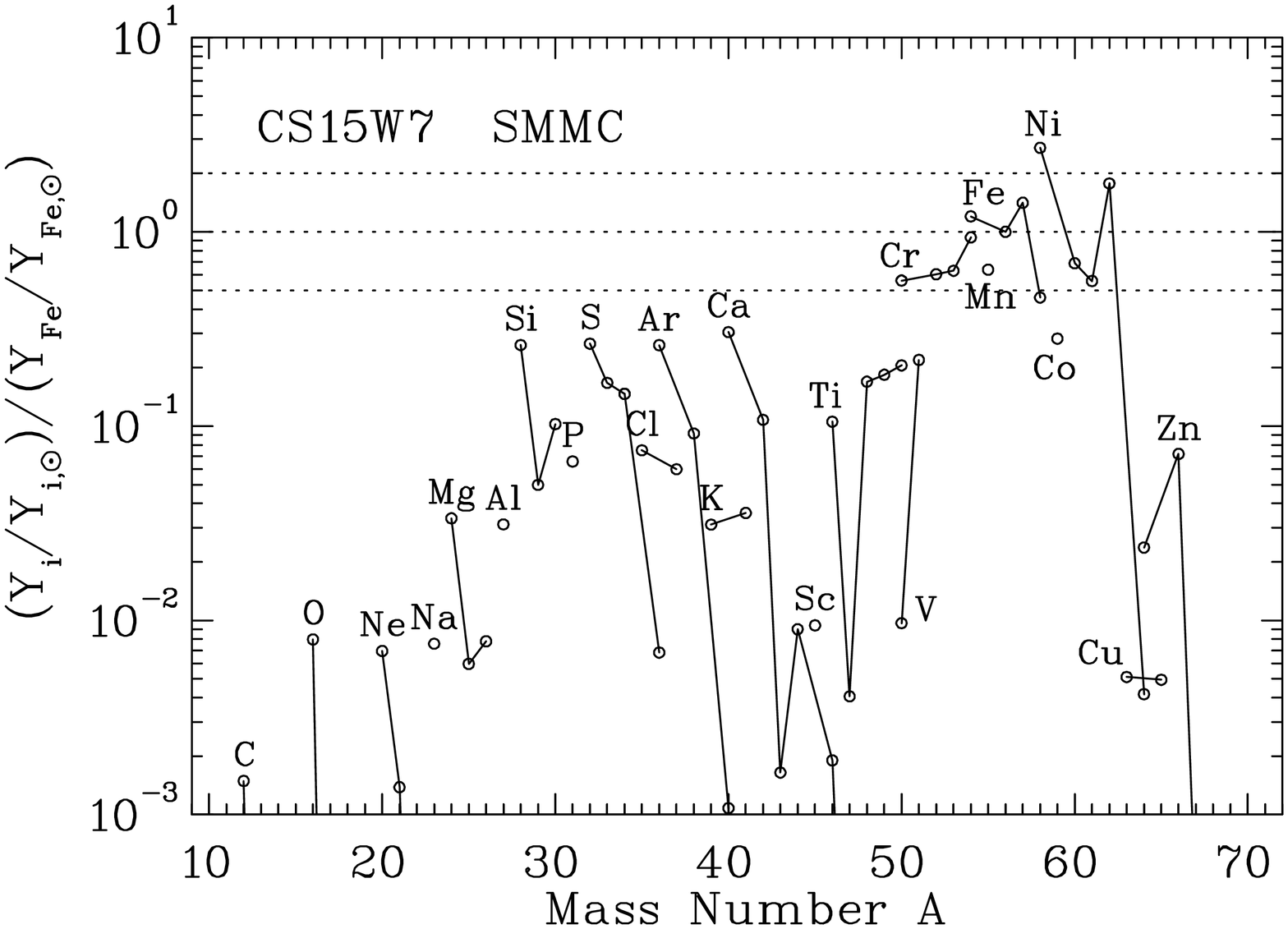}
    \caption{Similar to Figure 5 but for model CS15. The slight overproduction
    of $^{54}$Cr is reduced to solar ratios while $^{50}$Ti is even 
    underproduced by more than a factor of 3.
    \label{ysol3}}
\end{figure}

\begin{figure}[htp]
    \includegraphics[width=6cm,angle=90]{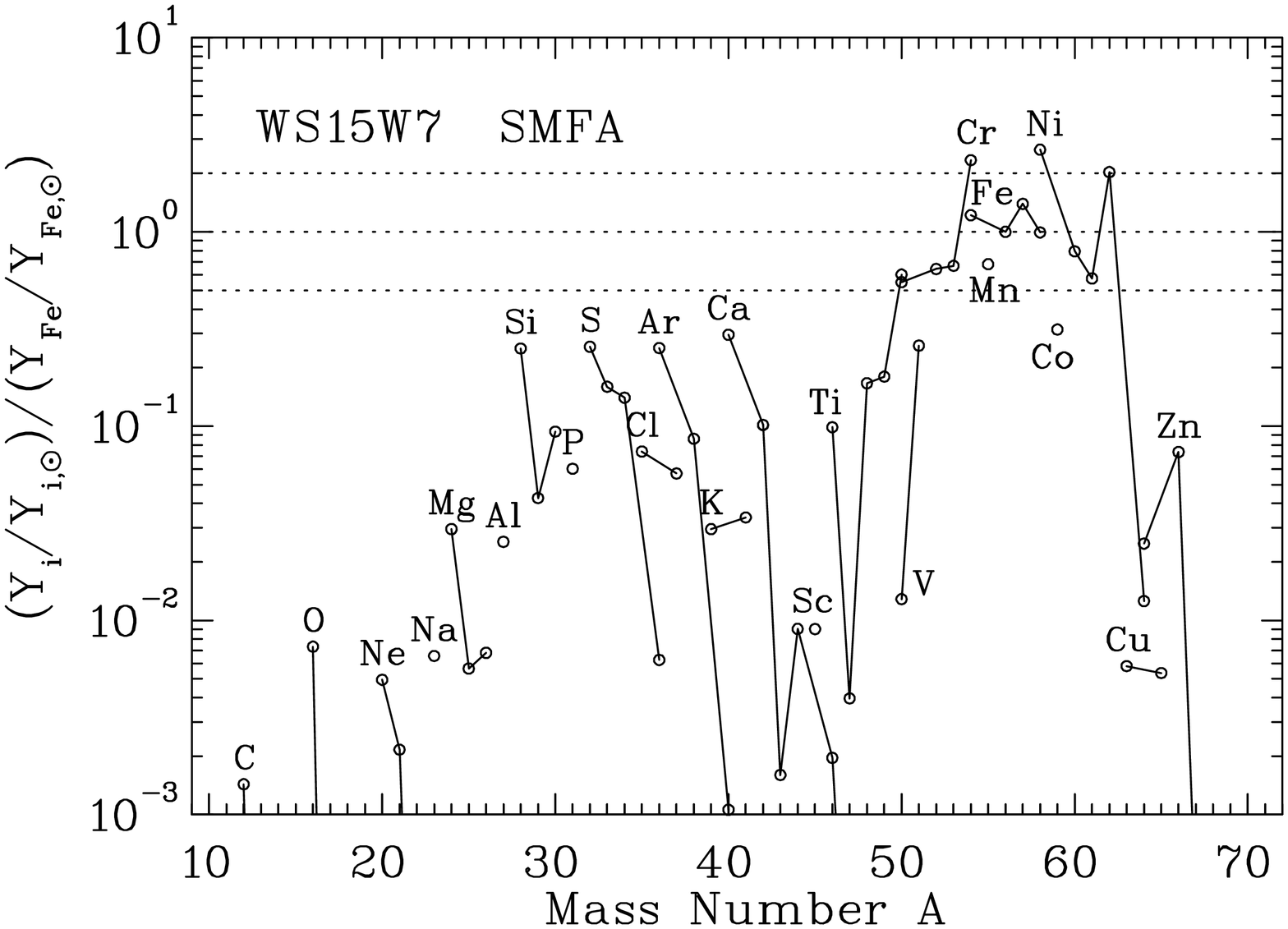}
    \includegraphics[width=6cm,angle=90]{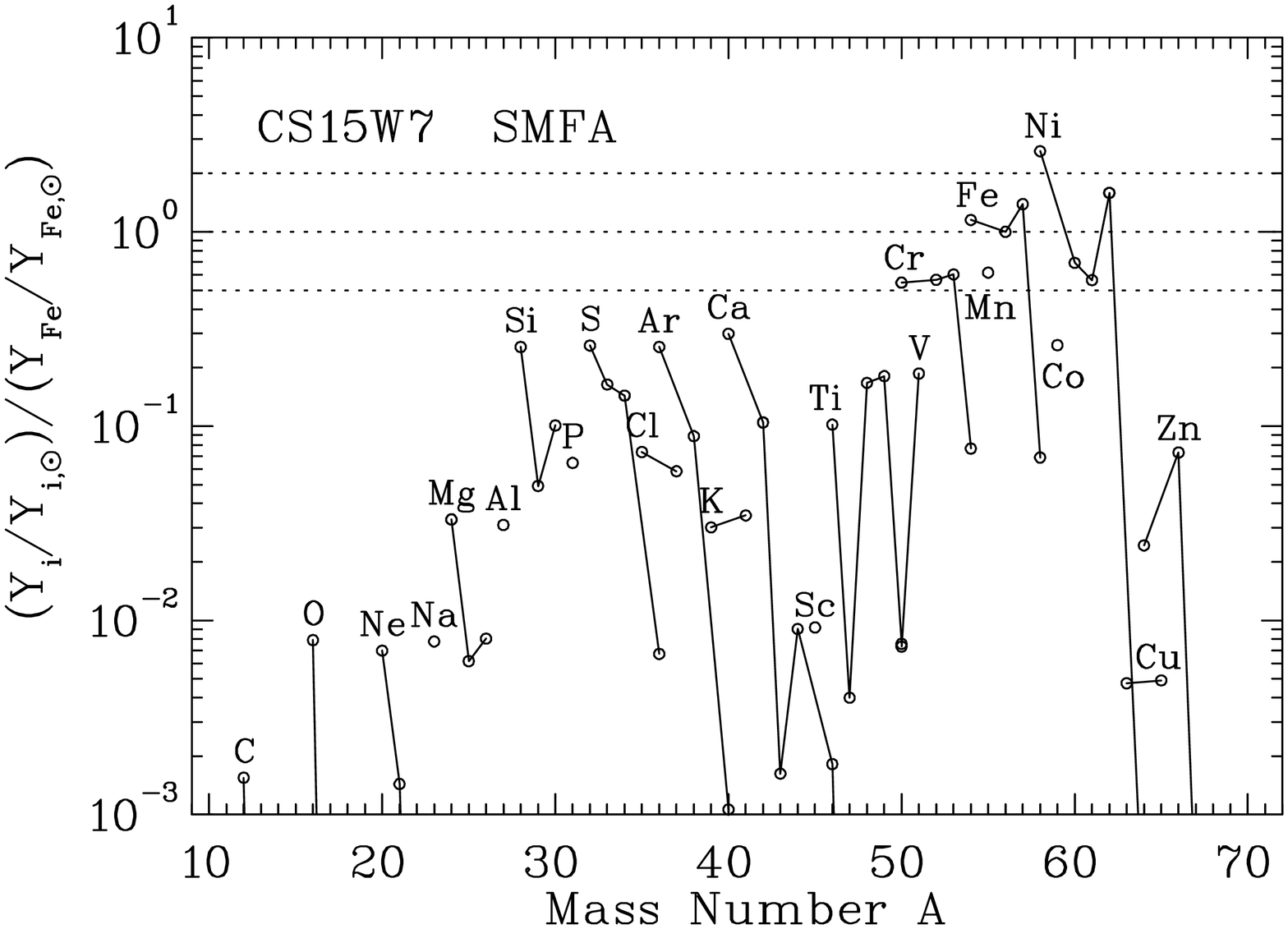}
    \caption{Similar to Figure 5 but for models WS15 and CS15, in both cases 
    with estimates of average factors (SMFA) from large-scale shell model 
    calculations for all nuclei. Even for WS15 the overproduction is reduced 
    close to the acceptable limit of a factor of 2. $^{50}$Ti and $^{54}$Cr 
    are now underproduced in CS15.
    \label{ysol2}}
\end{figure}

For SMMC the overabundances of the most neutron-rich nuclei
in WS15 is reduced to 75\% in comparison to the FFN calculations.
$^{50}$Ti and $^{54}$Cr are still significantly overabundant though. 
$^{58}$Fe is now reduced to about the limit of a factor 2. All three nuclei 
originate from the very center. The small overabundance of $^{58}$Ni and 
$^{62}$Ni remains (being due to the $Y_e$-gradient and thus $v_{def}$ rather
than $\rho_{ign}$ and the choice of electron capture rates).
In CS15 the abundance of $^{54}$Cr is reduced strongly. Again,
a slight overabundance of $^{58}$Ni is noticed.
Thus, the effect of the use of SMMC over FFN is the same in both models,
while the difference between WS15 and CS15 remains that the former is more
neutron-rich in the center, due to a higher $\rho_{ign}$).
For SMFA the overabundance of the neutron-rich nuclei is reduced to about 
20\% of the results obtained with FFN. 
In model WS15 the nuclei $^{50}$Ti,$^{58}$Fe, and $^{62}$Ni are now produced 
within a factor of 2 of solar values. Only a very slight overproduction of 
$^{54}$Cr and $^{58}$Ni remains, the latter being mostly dependent on the
$Y_e$-gradient (due to $v_{def}$) and unaffected by the rate change. 

$^{50}$Ti,$^{54}$Cr, and $^{58}$Fe, which are the dominant species
in matter with a $Y_{e}$-value below 0.452, are strongly 
reduced in WS15 for the rate set SMFA, because the central 
$Y_{e,c}$ of 0.451 is larger than 0.440 found using FFN. The number of mass 
zones with $Y_{e}$ below 0.452 is therefore
much smaller. This radial shift and central cut-off of the $Y_e$-curve and its 
result on abundances have already been discussed in relation to Figure 4 in 
section 3.1. The nuclei $^{58}$Ni and $^{54}$Fe are the dominant products
for $Y_{e}$ between 0.470 and 0.485, which extends over a very similar
range of integrated masses for all rate sets. The same is true for $^{56}$Fe, 
which is produced for $Y_{e}$ values between 0.46 and 0.47. The total
$^{62}$Ni abundance stems partially from $^{62}$Ni synthezised in the very 
center but is also partly due to the decay of $^{62}$Zn in the alpha-rich 
freeze-out zones of the outer core layers whose $Y_e$ is dominated by 
metallicity rather than
electron capture \citep{thinomyok86,thielemann97,iwamoto99}.
Therefore, the reduction of the abundance of this nucleus in the central sites 
does not affect the total value as much as is
the case of $^{54}$Cr and $^{50}$Ti, which both originate from central regions.
 
In comparison to the previous calculations employing FFN rates,
the model WS15 experiences strong improvement (i.e. reduction) in the 
overproduction of 
neutron-rich nuclei when applying the new sets of electron capture rates.
Model CS15, which showed no significant overproduction for 
the neutron-rich nuclei with FFN rates, still exhibits this same behavior.
In fact, the reduced electron capture rates cause a strong underproduction 
of neutron-rich nuclei like $^{50}$Ti and $^{54}$Cr in CS15.
Thus, the modified electron capture rates change the outcome of SN Ia models.
They certainly permit the higher ignition densities of model W ($2.1\times
10^9$ g cm$^{-3}$). If some of these neutron-rich nuclei originate only from 
SNe Ia, a shift for the average SN Ia close to this higher $\rho_{ign}$ of 
model W might even be needed. We have not yet addressed a possible upper
limit for  $\rho_{ign}$ when utilizing the present SMFA rate set.
A rough estimate can be obtained from Figure 8, where we show the central
$Y_e$ as a function of  $\rho_{ign}$ obtained with different electron
capture rate sets, making use of models CS15 
($\rho_{ign}=1.7\times 10^9$ g cm$^{-3}$) and WS15
($\rho_{ign}=2.1\times 10^9$ g cm$^{-3}$). If the trend continues in a 
similar way as experienced between models C and W, we would expect 
a central $Y_e$-value comparable to that of WS15 with FFN rates for
$\rho_{ign}=2.6\times 10^9$ g cm$^{-3}$ when utilizing SMFA. This corresponds
to an increased ignition density by about a factor of 1.24 when shifting from
FFN to SMFA rates.

\begin{figure}[tp]
\begin{center}
    \includegraphics[width=7cm,angle=90]{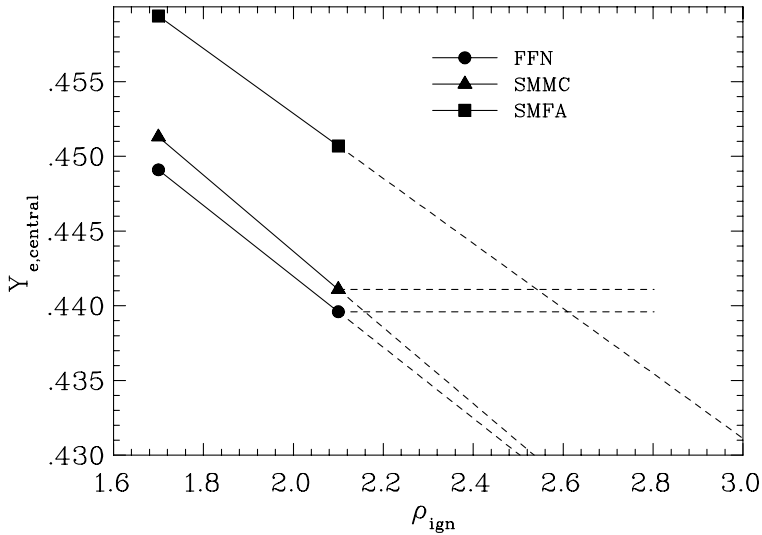}
    \caption{Change of central $Y_e$-values obtained for models CS15
    and WS15 (with increasing ignition densities $\rho_{ign}$ in units of
    $10^9$ g cm$^{-3}$) for different
    electron capture rate sets FFN, SMMC, and SMFA. If this trend continues
    to higher ignition densities in a similar way as the rough extrapolation
    indicated by dashed lines, the use of SMFA would lead to the same
    $Y_e$-value as WS15 with FFN for an ignition density of 
    $\rho_{ign}=2.6\times 10^9$ g cm$^{-3}$. This corresponds to an increase
    of a factor of 1.24 in $\rho_{ign}$.}
    \label{yedens}
\end{center}
\end{figure}

\section{Summary}
The need for an improved theoretical description of electron capture rates
for pf-shell nuclei beyond the early phenomenological tabulations by
\citet{ffn80,ffn82,ffn85} was realized by
\citet{Aufderheide91} and \citet{Aufderheide93a,Aufderheide93b}, 
calling for large shell model calculations which account for all correlations
among the valence nucleons in a major oscillator shell.
Motivated by the successful reproduction of all experimental GT$_+$ strength
distributions, the Shell Model Monte Carlo approach (SMMC) was recently  
applied to calculate stellar electron capture rates for several nuclei in the 
mass range $A=50-64$ \citep{dean98}.
The resulting significant modifications of stellar electron capture rates 
motivated the present investigation into potential effects on the dynamics and 
the nucleosynthesis of SNe Ia. 

The SMMC approach has, however, some limitations and requires supplemental
input. It makes use of a continuous strength distribution while the capture
rates on odd-$A$ nuclei should be supplemented by the contributions to
low-lying states. In addition,  rates for odd-odd nuclei are currently not
available. Thus, although SMMC calculations are generally substantiated by
large-scale shell model diagonalization calculations, information concerning
this low lying GT strength requires presently the use of the latter approach.
These approaches have recently made significant progress \citep{caur99a}.  For a
preliminary and approximate treatment we  followed the procedure outlined by
\citet{Martinez99}, who suggested to multiply the FFN electron capture rates
with averaged factors for even-even, odd-A, and odd-odd nuclei.

These electron capture rate modifications affect the early burning stage of a 
thermonuclear SN Ia explosion, being mostly responsible for the formation of
neutron-rich nuclei such as $^{48}$Ca,$^{50}$Ti, $^{54}$Cr, $^{54,58}$Fe, and 
$^{58}$Ni in the innermost zones of the SN Ia models.  
Our nucleosynthesis calculations show that, when the rates are multiplied with 
nuclear abundances, the odd-odd and odd-A nuclei cause the largest contribution
to the neutronization of nucleosynthesis ejecta and are essentially 
responsible for the $Y_{e}$-change, while the contribution of even-even nuclei 
is negligible. The present investigation focuses on the question whether our 
earlier conclusions
drawn for model parameters like central ignition density  $\rho_{ign}$, 
speed of the (deflagration) burning front  $v_{def}$, and the transition
density from deflagrations to detonations $\rho_{tr}$ 
\citep{iwamoto99}, based on FFN electron capture rates, would be
affected.

The transition density is always of the order $10^7$g cm$^{-3}$, where
electron capture rates are too slow on the dynamical timescales involved.
Thus, earlier conclusions drawn
for this parameter are unaffected. We found in the present analysis that
the $Y_e$-gradient is only determined by $v_{def}$ 
and apparently does not change with the set of electron capture rates.
Therefore, the conclusions for the permitted range of burning front
velocities also remain the same.
In \citet{iwamoto99} we determined this
range $v_{def}/v_s$ to be of the order 0.015-0.03. 

The central neutronization, however,
is dependent on $\rho_{ign}$ and - as shown here - on the
rate set of electron captures employed.
Thus the modified electron capture rates change the outcome of SN Ia yields.
In comparison to the previous calculations with FFN rates,
the model W ($\rho_{ign}$=$2.1\times 10^9$ g cm$^{-3}$) experienced a strong 
reduction in the overproduction of
neutron-rich nuclei when applying the new sets of electron capture rates.
In fact, they lie within the permitted uncertainties of solar Fe-group
abundances.
Model C ($\rho_{ign}$=$1.7\times 10^9$ g cm$^{-3}$), which showed no 
significant overproduction for the neutron-rich nuclei with FFN rates, 
still exhibits the same behavior.
In fact, the reduced electron captures rates cause a strong underproduction
of neutron-rich nuclei like $^{50}$Ti and $^{54}$Cr.
The rate modifications thus permit higher ignitions densities than previously
expected, by about a factor of 1.24. If some of these nuclei originate only 
from SNe Ia, the a shift to these higher $\rho_{ign}$ might be
needed for the average SN Ia. 

This work should, however, be completed using a
full set of shell model weak interaction rates.
In addition, strong Coulomb coupling between ions and electrons lowers the
electron capture Q-values and thus the threshold densities
\citep{couch73,bravo99}. Such a behavior, which is similar to the
screening of charged particle capture rates, has 
not yet been taken into account in nucleosynthesis
studies, but its importance should be tested. Some of the conclusions drawn 
here could be reversed, as its effect would cause an enhancement of
electron capture rates.

\acknowledgments
This work has been supported in part by 
the Swiss Nationalfonds (2000-53798.98), the US Department of Energy
(DOE contracts DE-AC05-96OR22464 and DE-FG02-96ER40983), the Danish Research 
Council, the grant-in-Aid for COE research (07CE2002)
of the Ministry of Education, Science, and Culture in Japan, a
fellowship of the Japan Society for the Promotion of Science for
Japanese Junior Scientists (6728), 
Some of us (KN and FKT) thank the Aspen Center for Physics 
for hospitality and inspiration during the 1999 Type Ia supernova program.

\end{document}